\documentclass{article}

\usepackage{natbib,amsthm}
\usepackage{amsmath,amssymb,amsfonts,bbm,verbatim,multirow}
\usepackage{epic,eepic,psfrag,epsfig}
\usepackage{subfigure}

\usepackage[dvips, plainpages = false, pdfpagelabels, 
                 bookmarks,
                 bookmarksopen = true,
                 bookmarksnumbered = true,
                 breaklinks = true,
                 linktocpage,
                 linkcolor = blue,
                 urlcolor  = black,
                 citecolor = black,
                 anchorcolor = black,
                 ]{hyperref}
\makeatletter
\def\url@mlcstyle{%
 \@ifundefined{selectfont}{\def\UrlFont{\sf}}{\def\UrlFont{\small\ttfamily}}}
\makeatother
\newtheorem{thm}{Theorem}

\newtheorem{prop}[thm]{Proposition}

\numberwithin{equation}{section}

\newenvironment{prooftitle}[1]{{\noindent \textsc{Proof #1}}\\}

\newenvironment{completionprooftitle}[1]{{\noindent 
\textsc{Completion of the Proof #1}}\\}

\title{Maximum likelihood estimation
  of a multidimensional log-concave density}

\author{Madeleine Cule \&  Richard Samworth\footnote{\emph{Address for correspondence}: 
Richard Samworth, Statistical Laboratory, Centre for Mathematical
Sciences, Wilberforce Road, Cambridge, UK. CB3 0WB. 
\newline E-mail: r.j.samworth@statslab.cam.ac.uk.}\\
  Statistical Laboratory\\
  University of Cambridge\\
  \and Michael Stewart \\
School of Mathematics and Statistics \\
University of Sydney
}

\begin{document}

\maketitle

\begin{abstract}
\noindent
  Let $X_1,\ldots,X_n$ be independent and identically distributed
  random vectors with a log-concave (Lebesgue) density $f$.  We first
  prove that, with probability one, there exists a unique maximum
  likelihood estimator $\hat{f}_n$ of $f$.  The use of this estimator
  is attractive because, unlike kernel density estimation, the method
  is fully automatic, with no smoothing parameters to choose.
  Although the existence proof is non-constructive, we are able to
  reformulate the issue of computing $\hat{f}_n$ in terms of a
  non-differentiable convex optimisation problem, and thus combine
  techniques of computational geometry with Shor's
  \mbox{$r$-algorithm} to produce a sequence that converges to
  $\hat{f}_n$.  For the moderate or large sample sizes in our
  simulations, the maximum likelihood estimator is shown to provide an
  improvement in performance compared with kernel-based methods, even
  when we allow the use of a theoretical, optimal fixed bandwidth for
  the kernel estimator that would not be available in practice.  We
  also present a real data clustering example, which shows that our
  methodology can be used in conjunction with the
  Expectation--Maximisation (EM) algorithm to fit finite mixtures of
  log-concave densities.  An \texttt{R} version of the algorithm is
  available in the package \texttt{LogConcDEAD} -- Log-Concave Density
  Estimation in Arbitrary Dimensions.
  
\bigskip
\noindent {\bf Keywords:} Computational geometry, log-concavity,
maximum likelihood estimation, non-differentiable convex optimisation,
nonparametric density estimation, Shor's $r$-algorithm
\end{abstract}

\section{Introduction}
\label{Sec:Intro}

Modern nonparametric density estimation began with the introduction of
a kernel density estimator in the pioneering work of
\citet{FixHodges1951}, later republished as \citet{FixHodges1989}.
For independent and identically distributed real-valued observations,
the appealing asymptotic theory of the mean integrated squared error
was provided by \citet{Rosenblatt1956} and \citet{Parzen1962}.  This
theory leads to an asymptotically optimal choice of the smoothing
parameter, or bandwidth.  Unfortunately, however, it depends on the
unknown density $f$ through the integral of the square of the second
derivative of $f$.  Considerable effort has therefore been focused on
finding methods of automatic bandwidth selection (cf. Wand and Jones,
1995, Chapter 3, and the references therein).  Although this has
resulted in algorithms, e.g. \citet{Chiu1992}, that achieve the
optimal rate of convergence of the relative error, namely
$O_p(n^{-1/2})$, where $n$ is the sample size, good finite sample
performance is by no means guaranteed.

This problem is compounded when the observations take values in
$\mathbb{R}^d$, where the general kernel estimator
\citep{Deheuvels1977} requires the specification of a symmetric,
positive definite $d \times d$ bandwidth matrix.  The difficulties
involved in making the $d(d+1)/2$ choices for its entries mean that
attention is often restricted either to bandwidth matrices that are
diagonal, or even to those that are scalar multiples of the identity
matrix.  Of course, practical issues of automatic bandwidth selection
remain.

In this paper, we propose a fully automatic nonparametric estimator of
$f$, with no tuning parameters to be chosen, under the condition that
$f$ is log-concave -- that is, $\log f$ is a concave function.  The
class of log-concave densities has many attractive properties and has
been well-studied, particularly in the economics, sampling and
reliability theory literature.  See Section~\ref{Sec:LCproperties} for
further discussion of examples, applications and properties of
log-concave densities.

\begin{figure}
\begin{minipage}[b]{0.47\linewidth} 
\centering
\includegraphics[width=2.2in,height=2.4in]{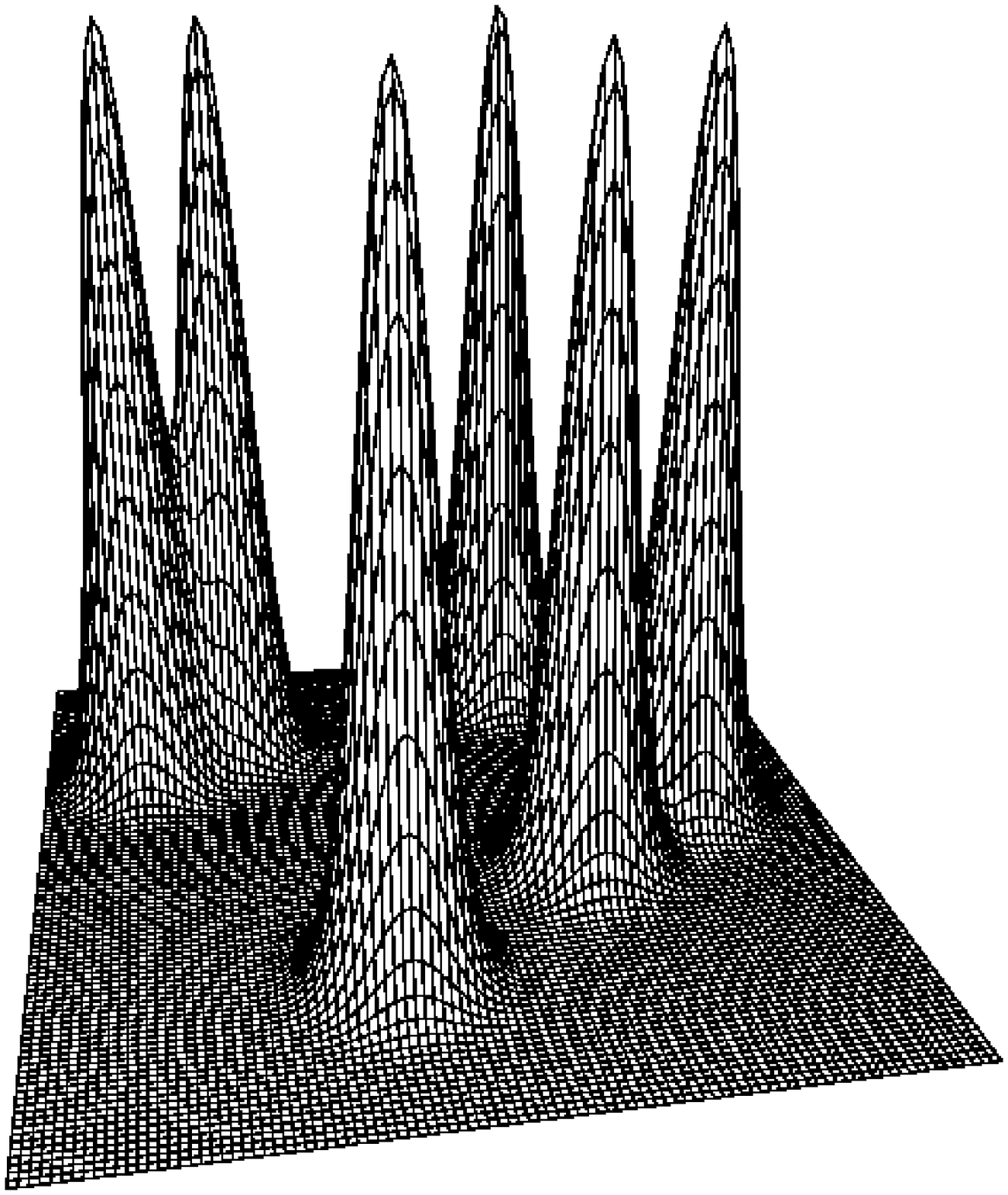}
\caption{\label{Fig:Unbounded}Without any shape
constraint on the class of densities, the likelihood function is
unbounded, because we can take successively close approximations to a
mixture of $n$ `spikes' (one on each $X_i$).}
\end{minipage}
\hspace{0.05\linewidth}
\begin{minipage}[b]{0.47\linewidth}
\centering
\includegraphics[width=2.2in,height=2.2in]{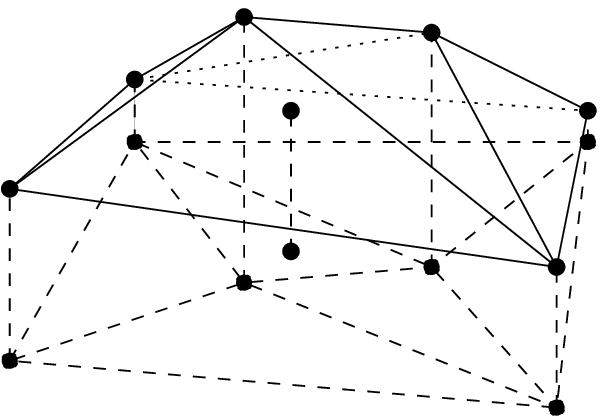}
\caption{\label{Fig:Tent}  The
`tent-like' structure of the graph of the logarithm of the maximum
likelihood estimator for bivariate data.}
\end{minipage}
\end{figure}

In Section~\ref{Sec:ExistUnique}, we show that if $X_1,\ldots,X_n$ are
independent and identically distributed random vectors with a
log-concave density, then with probability one there exists a unique
log-concave density $\hat{f}_n$ that maximises the likelihood
function,
\[ L(f) = \prod_{i=1}^n f(X_i).
\] Before continuing, it is worth noting that without any shape
constraints on the densities under consideration, the likelihood
function is unbounded.  To see this, we could define a sequence
$(f_n)$ of densities that represent successively close approximations
to a mixture of $n$ `spikes' (one on each $X_i$), such as $f_n(x) =
n^{-1}\sum_{i=1}^n \phi_{d,n^{-1}I}(x-X_i)$, where $\phi_{d,\Sigma}$
denotes the $N_d(0,\Sigma)$ density.  This sequence satisfies $L(f_n)
\rightarrow \infty$ as $n \rightarrow \infty$
(cf.~Figure~\ref{Fig:Unbounded}).  In fact, a modification of this
argument may be used to show that the likelihood function remains
unbounded even if we restrict attention to unimodal densities.

\begin{figure}
\centering
\subfigure[Density]{ 
\includegraphics[width=2.5in,height=2in]{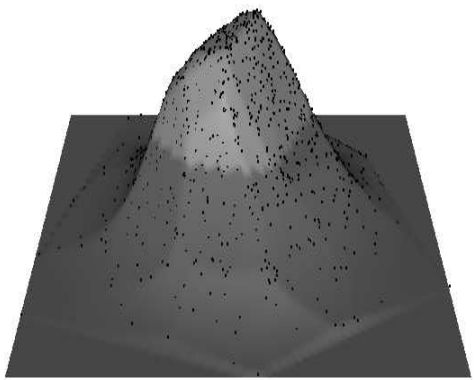}
}
\hspace{0.5in} 
\subfigure[Log-density]{
\includegraphics[width=2.5in,height=2in]{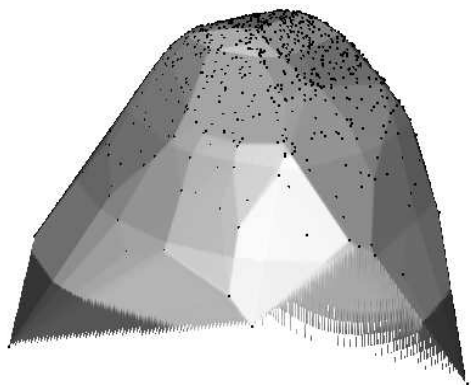}
}
\caption{\label{Fig:Example}Log-concave maximum likelihood estimates based on 
1000 observations (plotted as dots) from a standard bivariate normal
distribution.}
\end{figure}

Figure~\ref{Fig:Tent} gives a diagram illustrating the structure
of the maximum likelihood estimator on the logarithmic scale.  This
structure is most easily visualised for two-dimensional data, where
one can imagine associating a `tent pole' with each observation,
extending vertically out of the plane.  For certain tent pole heights,
the graph of the logarithm of the maximum likelihood estimator can be
thought of as the roof of a taut tent stretched over the tent poles.
The fact that the logarithm of the maximum likelihood estimator is of
this `tent function' form constitutes part of the proof of its
existence and uniqueness.

In Section~\ref{Sec:Computation}, we discuss the computational problem
of how to adjust the $n$ tent pole heights so that the corresponding
tent functions converge to the logarithm of the maximum likelihood
estimator.  One reason that this computational problem is so
challenging in more than one dimension is the fact that it is
difficult to describe the set of tent pole heights that correspond to
concave functions.  The key observation, discussed in
Section~\ref{Sec:Computation}, is that it is possible to minimise a
modified objective function that it is convex (though
non-differentiable).  This allows us to apply the powerful
non-differentiable convex optimisation methodology of the subgradient
method~\citep{Shor1985} and a variant called Shor's $r$-algorithm,
which has been implemented by~\citet{KappelKuntsevich2000}.

As an illustration of the estimates obtained, Figure \ref{Fig:Example}
presents plots of the maximum likelihood estimator, and its logarithm,
for 1000 observations from a standard bivariate normal distribution.
These plots were created using the \texttt{LogConcDEAD} package
\citep{CGS2007} in \texttt{R} \citep{R}, which exploits the interactive
surface-plotting software available in the \texttt{rgl} package
\citep{AdlerMurdoch2007}.

In Section~\ref{Sec:Simulations} we present simulations to compare the
finite-sample performance of the maximum likelihood estimator with
kernel-based methods.  The results are striking: even when we use the
theoretical, optimal bandwidth for the kernel estimator (or an
asymptotic approximation to this when it is not available), we find
that the maximum likelihood estimator has a rather smaller mean
integrated squared error for moderate or large sample sizes, despite
the fact that this optimal bandwidth depends on properties of the
density that would be unknown in practice.  This suggests that the 
maximum likelihood estimator is able to adapt to the local smoothness 
of the underlying density automatically.

Nonparametric density estimation is a fundamental tool for the
visualisation of structure in exploratory data analysis, and has an
enormous literature that includes the monographs of
\citet{DevroyeGyorfi1985}, \citet{Silverman1986}, \citet{Scott1992}
and \citet{WandJones1995}.  Our proposed method may certainly be used
for this purpose; however, it may also be used as an intermediary
stage in more involved statistical procedures.  For instance:
\begin{enumerate}
\item In classification problems, we have $p \geq 2$ populations of
interest, and assume in this discussion that these have densities
$f_1,\ldots,f_p$ on $\mathbb{R}^d$.  We observe training data of the
form $\{(X_i,Y_i):i=1,\ldots,n\}$, where if $Y_i = j$, then $X_i$ has
density $f_j$.  The aim is to classify a new observation $z \in
\mathbb{R}^d$ as coming from one of the populations.  Problems of this
type occur in a huge variety of applications, including medical
diagnosis, archaeology, ecology etc. -- see \citet{Gordon1981},
\citet{Hand1981} or \citet{DGL1996} for further details and examples.
A natural approach to classification problems is to construct density
estimates $\hat{f}_1,\ldots,\hat{f}_p$, where $\hat{f}_j$ is based on
the $n_j$ observations, say, from the $j$th population, namely $\{X_i:
Y_i = j\}$.  We may then assign $z$ to the $j$th population if
$n_j\hat{f}_j(z) = \max\{n_1\hat{f}_1(z),\ldots,n_p\hat{f}_p(z)\}$.
In this context, the use of kernel-based estimators in general
requires the choice of $p$ separate $d \times d$ bandwidth matrices,
while the corresponding procedure based on the log-concave maximum
likelihood estimates is again fully automatic.
\item Clustering problems are closely related to the classification
problems described above.  The difference is that, in the above
notation, we do not observe $Y_1,\ldots,Y_n$, and have to assign each
of $X_1,\ldots,X_n$ to one of the $p$ populations.  A common technique
is based on fitting a mixture density of the form $f(x) = \sum_{j=1}^p
\pi_j f_j(x)$, where the mixture proportions $\pi_1,\ldots,\pi_p$ are
positive and sum to one.  Under the assumption that each of the
component densities $f_1,\ldots,f_p$ is log-concave, we show in
Section~\ref{Sec:Clustering} that our methodology can be extended to
fit such a finite mixture density, which need not itself be
log-concave -- cf.~Section~\ref{Sec:LCproperties}.  We also illustrate
this clustering algorithm on a Wisconsin breast cancer data set in
Section~\ref{Sec:Clustering}, where the aim is to separate
observations into benign and malignant component populations.
\item A functional of the true underlying density may be estimated by
the corresponding functional of a density estimator, such as the
log-concave maximum likelihood estimator.  Examples of functionals of
interest include probabilities, such as $\int_{\|x\| \geq 1} f(x) \,
dx$, moments, e.g. $\int \|x\|^2 f(x) \, dx$, and the differential
entropy, $-\int f(x) \log f(x) \, dx$.  It may be possible to compute
the plug-in estimator based on the log-concave maximum likelihood
estimator analytically, but in Section~\ref{Sec:Sample}, we show that
even if this is not possible, in many cases of interest we can sample
from the log-concave maximum likelihood estimator $\hat{f}_n$, and
hence obtain a Monte Carlo estimate of the functional.  This nice
feature also means that the log-concave maximum likelihood estimator
can be used in a Monte Carlo bootstrap procedure for assessing
uncertainty in functional estimates -- see Section~\ref{Sec:Sample}
for further details.
\item The fitting of a nonparametric density estimate may give an
indication of the validity of a particular smaller model (often
parametric).  Thus, a contour plot of the log-concave maximum
likelihood estimator may provide evidence that the underlying density
has elliptical contours, and thus suggest that a model that exploits
this elliptical symmetry.
\item In the univariate case, \citet{Walther2002} describes
methodology based on log-concave density estimation for addressing the
problem of detecting the presence of mixing in a distribution.  As an
application, he cites the Pickering/Platt debate \citep{Swales1985} on
the issue of whether high blood pressure is a disease (in which case
observed blood pressure measurements should follow a mixture
distribution), or simply a label attached to people in the right tail
of the blood pressure distribution.  As a result of our algorithm for
computing the multidimensional log-concave maximum likelihood
estimator, this methodology extends immediately to more than one
dimension.
\end{enumerate}
   
There has been considerable recent interest in shape-restricted
nonparametric density estimation, but most of it has been confined to
the case of univariate densities, where the computational algorithms
are more straightforward.  Nevertheless, as was discussed above, it is
in multivariate situations that the automatic nature of the maximum
likelihood estimator is particularly valuable.  \citet{Walther2002},
\citet{DumbgenRufibach2007} and \citet{PWM2007} have proved the
existence and uniqueness of the log-concave maximum likelihood
estimator in one dimension and \citet{DumbgenRufibach2007},
\citet{PWM2007} and \citet{BRW2008} have studied its theoretical
properties.  \citet{Rufibach2007} has compared different algorithms
for computing the univariate estimator, including the iterative convex
minorant algorithm~\citep{GroeneboomWellner1992,Jongbloed1998}, and
three others.  \citet{DHR2007} also present an Active Set algorithm,
which has similarities with the vertex direction and vertex reduction
algorithms described in \citet{GJW2008}.  For univariate data, it is
also well-known that there exist maximum likelihood estimators of a
non-increasing density supported on $[0,\infty)$~\citep{Grenander1956}
and of a convex, decreasing density~\citep{GJW2001}.

In Section~\ref{Sec:Discussion}, we give a brief concluding
discussion, and suggest some directions for future research.  Finally,
we present in Appendix~\ref{Sec:Glossary} a glossary of terms and
results from convex analysis and computational geometry that appear in
italics at their first occurrence in the main body of the paper; the
references are~\citet{Rockafellar1997} and~\citet{Lee1997}.  Proofs
are deferred to Appendix~\ref{Sec:Proofs}, except that the beginning of the proof
of Theorem~\ref{Thm:ExistUnique} is given in the main text, as the
ideas and notation introduced are needed in the remainder of the
paper.

\section{Log-concave densities: examples, applications and properties}
\label{Sec:LCproperties}

Many of the most commonly-encountered parametric families of
univariate distributions have \emph{log-concave} densities, including
the family of normal distributions, gamma distributions with shape
parameter at least one, $\mathrm{Beta}(\alpha,\beta)$ distributions
with $\alpha, \beta \geq 1$, Weibull distributions with shape
parameter at least one, Gumbel, logistic and Laplace densities; see
\citet{BagnoliBergstrom1989} for other examples.  Univariate
log-concave densities are unimodal and have fairly light tails -- it
may help to think of the exponential distribution (where the logarithm
of the density is a linear function on the positive half-axis) as a
borderline case.  Thus Cauchy, Pareto and lognormal densities, for
instance, are not log-concave.  Mixtures of log-concave densities may
be log-concave, but in general they are not; for instance, for $p \in
(0,1)$, the location mixture of standard univariate normal densities
$f(x) = p \phi(x) + (1-p)\phi(x-\mu)$ is log-concave if and only if
$\|\mu\| \leq 2$.

The assumption of log-concavity is a popular one in economics;
\citet{CaplinNaelbuff1991a} show that in the theory of elections and
under a log-concavity assumption, the proposal most preferred by the
mean voter is unbeatable under a 64\% majority rule.  As another
example, in the theory of imperfect competition,
\citet{CaplinNaelbuff1991b} use log-concavity of the density of
consumers' utility parameters as a sufficient condition in their proof
of the existence of a pure-strategy price equilibrium for any number
of firms producing any set of products.  See
\citet{BagnoliBergstrom1989} for many other applications of
log-concavity to economics.  \citet{Brooks1998}
and~\citet{MengersenTweedie1996} have exploited the properties of
log-concave densities in studying the convergence of Markov chain
Monte Carlo sampling procedures.

\citet{An1998} lists many useful properties of log-concave densities.
For instance, if $f$ and $g$ are (possibly multidimensional)
log-concave densities, then their convolution $f \ast g$ is
log-concave.  In other words, if $X$ and $Y$ are independent and have
log-concave densities, then their sum $X+Y$ has a log-concave density.
The class of log-concave densities is also closed under the taking of
pointwise limits.  One-dimensional log-concave densities have
increasing hazard functions, which is why they are of interest in
reliability theory.  Moreover, \citet{Ibragimov1956} proved the
following characterisation: a univariate density $f$ is log-concave if
and only if the convolution $f \ast g$ is unimodal for every unimodal
density $g$.  There is no natural generalisation of this result to
higher dimensions.

As was mentioned in Section~\ref{Sec:Intro}, this paper concerns
multidimensional log-concave densities, for which fewer properties are
known.  It is therefore of interest to understand how the property of
log-concavity in more than one dimension relates to the univariate
notion.  Our first proposition below is intended to give some insight
into this issue.  It is not formally required for the subsequent
development of our methodology in Sections~\ref{Sec:ExistUnique}
and~\ref{Sec:Computation}, although we did apply the result when
designing our simulation study in Section~\ref{Sec:Simulations}.  We
assume throughout that log-concave densities are with respect to
Lebesgue measure on the \emph{affine hull} of their \emph{support},
and `$X$ has a log-concave density' means `there exists a version of
the density of $X$ that is log-concave'.
\begin{prop}
\label{Prop:LCProperties} Let $X$ be a $d$-variate random vector
having density $f$ with respect to Lebesgue measure on $\mathbb{R}^d$.
For a subspace $V$ of $\mathbb{R}^d$, let $P_V(x)$ denote the
orthogonal projection of $x$ onto $V$.  Then in order that $f$ be
log-concave, it is:
\begin{enumerate}
\item necessary that for any subspace $V$, the marginal density of
$P_V(X)$ is log-concave and the conditional density
$f_{X|P_V(X)}(\cdot|t)$ of $X$ given $P_V(X)=t$ is log-concave for
each $t$
\item sufficient that for every $(d-1)$-dimensional subspace $V$, the
conditional density $f_{X|P_V(X)}(\cdot|t)$ of $X$ given $P_V(X)=t$ is
log-concave for each $t$.
\end{enumerate}
\end{prop} The part of Proposition~\ref{Prop:LCProperties}(a)
concerning marginal densities is an immediate consequence of Theorem~6
of~\citet{Prekopa1973}.  One can regard
Proposition~\ref{Prop:LCProperties}(b) as saying that a
multidimensional density is log-concave if the restriction of the
density to any line is a (univariate) log-concave function.

It is interesting to compare the properties of log-concave densities
presented in Proposition~\ref{Prop:LCProperties} with the
corresponding properties of Gaussian densities.  In fact,
Proposition~\ref{Prop:LCProperties} remains true if we replace
`log-concave' with `Gaussian' throughout (at least, provided that in
part~(b) we also assume there is a point at which $f$ is twice
differentiable).  These shared properties suggest that the class of
log-concave densities is a natural, infinite-dimensional
generalisation of the class of Gaussian densities.

\section{Existence, uniqueness and structure of the maximum likelihood
estimator}
\label{Sec:ExistUnique}

Let $\mathcal{F}_0$ denote the class of log-concave densities on
$\mathbb{R}^d$ with $d$-\emph{dimensional} support, and let $f_0 \in
\mathcal{F}_0$.  The degenerate case where the support is of dimension
smaller than $d$ can also be handled, but for simplicity of exposition
we concentrate on the non-degenerate case.  Suppose that
$X_1,\ldots,X_n$ are a random sample from $f_0$.  We say that
$\hat{f}_n = \hat{f}_n(X_1,\ldots,X_n) \in \mathcal{F}_0$ is a
(nonparametric) \emph{maximum likelihood estimator} of $f_0$ if it
maximises $\ell(f) = \sum_{i=1}^n \log f(X_i)$ over $f \in
\mathcal{F}_0$.

\begin{thm}
\label{Thm:ExistUnique} Suppose that $n \geq d+1$.  Then, with
probability one, a nonparametric maximum likelihood estimator
$\hat{f}_n$ of $f_0$ exists and is unique.
\end{thm} \textsc{First Part of Proof}.  We may assume that
$X_1,\ldots,X_n$ are distinct and their convex hull, $C_n =
\mathrm{conv}(X_1,\ldots,X_n)$, is a $d$-dimensional \emph{polytope}
(an event of probability one when $n \geq d+1$).  By a standard
argument in convex analysis \citep[][p.~37]{Rockafellar1997}, for each
$y = (y_1,\ldots,y_n) \in \mathbb{R}^n$ there exists a function
$\bar{h}_y:\mathbb{R}^d \rightarrow \mathbb{R}$ with the property that
$\bar{h}_y$ is the least concave function satisfying $\bar{h}_y(X_i)
\geq y_i$ for all $i=1,\ldots,n$.  Informally, $\bar{h}_y$ is a `tent
function', and a typical example is depicted in
Figure~\ref{Fig:Tent}.  Let $\mathcal{H} = \{\bar{h}_y: y \in
\mathbb{R}^n\}$ denote `the class of tent functions'.  Let
$\mathcal{F}$ denote the set of all log-concave functions on
$\mathbb{R}^d$, and for $f \in \mathcal{F}$, define
\[ \psi_n(f) = \frac{1}{n}\sum_{i=1}^n \log f(X_i) -
\int_{\mathbb{R}^d} f(x) \, dx.
\] Suppose that $f$ maximises $\psi_n(\cdot)$ over $\mathcal{F}$.  The
main part of the proof, which is completed in the Appendix, consists
of showing that \renewcommand{\labelenumi}{(\theenumi)}
\renewcommand{\theenumi}{\roman{enumi}}
\begin{enumerate}
\item $f(x) > 0$ for $x \in C_n$
\item $f(x) = 0$ for $x \notin C_n$
\item $\log f \in \mathcal{H}$
\item $f \in \mathcal{F}_0$
\item there exists $M > 0$ such that if $\max_i |\bar{h}_y(X_i)| \geq
M$, then $\psi_n\bigl(\exp(\bar{h}_y)\bigr) \leq \psi_n(f)$.
\end{enumerate}

Although step~(iii) above gives us a finite-dimensional class of
functions to which $\log \hat{f}_n$ belongs, the proof of
Theorem~\ref{Thm:ExistUnique} gives no indication of how to find the
member of this class that maximises the likelihood function.  We
therefore seek an iterative algorithm to compute the estimator, but
first we describe the structure we see in Figure~\ref{Fig:Tent} in 
Section~\ref{Sec:Intro} more precisely.  From now on, we assume:
\begin{description}
\item[(\textbf{A1}):] $n \geq d+1$, and every subset of
$\{X_1,\ldots,X_n\}$ of size $d+1$ is \emph{affinely independent}.
\end{description} Note that when $n \geq d+1$, the event in
\textbf{(A1)} has probability one.  From step~(iii) in the proof of
Theorem~\ref{Thm:ExistUnique} above, there exists $y \in \mathbb{R}^n$
such that $\log \hat{f}_n = \bar{h}_y$.  As illustrated in
Figure~\ref{Fig:Tent}, and justified
formally by Corollary~17.1.3 and Corollary 19.1.2 of
\citet{Rockafellar1997}, the convex hull of the data, $C_n$, may be
\emph{triangulated} in such a way that $\log \hat{f}_n$ coincides with an
\emph{affine function} on each \emph{simplex} in the triangulation.
In other words, if $j = (j_1,\ldots,j_{d+1})$ is a $(d+1)$-tuple of
distinct indices in $\{1,\ldots,n\}$, and $C_{n,j} =
\mathrm{conv}(X_{j_1},\ldots,X_{j_{d+1}})$, then there exists a finite
set $J$ consisting of $m$ such $(d+1)$-tuples, with the following
three properties:
\begin{enumerate}
\item $\cup_{j \in J} C_{n,j} = C_n$
\item the \emph{relative interiors} of the sets $\{C_{n,j}:j \in J\}$
are pairwise disjoint
\item
\[ \log \hat{f}_n(x) = \left\{ \begin{array}{ll} \langle x,b_j \rangle -
\beta_j & \mbox{\text{if $x \in C_{n,j}$ for some $j \in J$}} \\
-\infty & \mbox{if $x \notin C_n$} \end{array} \right.
\] for some $b_1,\ldots,b_m \in \mathbb{R}^d$ and
$\beta_1,\ldots,\beta_m \in \mathbb{R}$.  Here and below, $\langle
\cdot, \cdot \rangle$ denotes the usual Euclidean inner product in
$\mathbb{R}^d$.
\end{enumerate} In the iterative algorithm that we propose in
Section~\ref{Sec:Computation} for computing the maximum likelihood
estimator, we need to find convex hulls and triangulations at each
iteration.  Fortunately, these can be computed efficiently using
the \texttt{Quickhull} algorithm of \citet{BDH1996}.

\section{Computation of the maximum likelihood estimator}
\label{Sec:Computation}

\subsection{Reformulation}
As a first attempt to find an algorithm which produces a sequence that
converges to the maximum likelihood estimator in
Theorem~\ref{Thm:ExistUnique}, it is natural to try to minimise
numerically the function
\[ \tau(y_1,\ldots,y_n) = -\frac{1}{n}\sum_{i=1}^n \bar{h}_y(X_i) +
\int_{C_n} \exp\{\bar{h}_y(x)\} \, dx.
\] Although this approach might work in principle, one difficulty is
that $\tau$ is not convex, so this approach is extremely
computationally intensive, even with relatively few observations.
Another reason for the numerical difficulties stems from the fact that
the set of $y$-values on which $\tau$ attains its minimum is rather
large: in general it may be possible to alter particular components
$y_i$ without changing $\bar{h}_y$.  Of course, we could have defined
$\tau$ as a function of $\bar{h}_y$ rather than as a function of the
vector of tent pole heights $y=(y_1,\ldots,y_n)$.  Our choice,
however, motivates the following definition of a modified objective
function:
\begin{equation}
\label{Eq:sigmadef} \sigma(y_1,\ldots,y_n) = -\frac{1}{n}\sum_{i=1}^n
y_i + \int_{C_n} \exp\{\bar{h}_y(x)\} \, dx.
\end{equation} The great advantages of minimising $\sigma$ rather than
$\tau$ are seen by the following theorem.
\begin{thm}
\label{Thm:sigma} Assume \textbf{(A1)}.  The function $\sigma$ is a
convex function satisfying $\sigma \geq \tau$.  It has a unique
minimum at $y^* \in \mathbb{R}^n$, say, and $\log \hat{f}_n =
\bar{h}_{y^*}$.
\end{thm} Thus Theorem~\ref{Thm:sigma} shows that the unique minimum
$y^* = (y_1^*,\ldots,y_n^*)$ of $\sigma$ belongs to the minimum set of
$\tau$.  In fact, it corresponds to the element of the minimum set for
which $\bar{h}_{y^*}(X_i) = y_i^*$ for $i=1,\ldots,n$.  Informally,
then, $\bar{h}_{y^*}$ is `a tent function with all of the tent poles
touching the tent'.

In order to compute the function $\sigma$ at a generic point $y =
(y_1,\ldots,y_n) \in \mathbb{R}^n$, we need to be able to evaluate the
integral in~(\ref{Eq:sigmadef}).  In the notation of
Section~\ref{Sec:ExistUnique}, we may write
\[ \int_{C_n} \exp\{\bar{h}_y(x)\} \, dx = \sum_{j \in J}
\int_{C_{n,j}} \exp\{ \langle x,b_j \rangle - \beta_j\} \, dx.
\] For each $j = (j_1,\ldots,j_{d+1}) \in J$, let $A_j$ be the $d
\times d$ matrix whose $l$th column is $X_{j_{l+1}} - X_{j_1}$ for
$l=1,\ldots,d$, and let $\alpha_j = X_{j_1}$.  Then the \emph{affine
transformation} $w \mapsto A_j w + \alpha_j$ takes the unit simplex
$T_d = \bigl\{w = (w_1,\ldots,w_d): w_l \geq 0, \sum_{l=1}^d w_l \leq
1\bigr\}$ to $C_{n,j}$.  Letting $z_{j,l} = y_{j_{l+1}} - y_{j_1}$, we
can then establish by a simple change of variables and induction on
$d$ that if $z_{j,1},\ldots,z_{j,d}$ are non-zero and distinct, then
\begin{equation}
\label{Eq:Int} \int_{C_n} \exp\{\bar{h}_y(x)\} \, dx = \sum_{j \in J}
|\det A_j|e^{y_{j_1}} \sum_{r=1}^d \frac{e^{z_{j,r}} - 1}{z_{j,r}}
\prod_{\stackrel{\scriptstyle{1 \leq s \leq d}}{s \neq
r}}\frac{1}{z_{j,r}-z_{j,s}}.
\end{equation} Further details
of this calculation can be found in a longer version of this paper
\citep{CSS2008}.  The singularities that occur when some of
$z_{j,1},\ldots,z_{j,d}$ may be zero or equal are removable.  Thus,
although~(\ref{Eq:Int}) is a little complicated, it allows the
computation of our objective function.

\subsection{Nonsmooth optimisation}
There is a vast literature on techniques of convex optimisation
(cf.~\citet{BoydVandenberghe2004}, for example), including the method
of steepest descent and Newton's method.  Unfortunately, these methods
rely on the differentiability of the objective function, and the
function $\sigma$ is not differentiable.  This can be seen informally
by studying the schematic diagram in Figure~\ref{Fig:Tent} again.  If
the $i$th tent pole, say, is touching but not critically supporting
the tent, then decreasing the height of this tent pole does not change
the tent function, and thus does not alter the integral
in~(\ref{Eq:sigmadef}); on the other hand, increasing the height of
the tent pole does alter the tent function and therefore the integral
in~(\ref{Eq:sigmadef}).  This argument may be used to show that at
such a point, the $i$th partial derivative of $\sigma$ does not exist.

The set of points at which $\sigma$ is not differentiable constitute a
set of Lebesgue measure zero, but the non-differentiability cannot be
ignored in our optimisation procedure.  Instead, it will be necessary
to derive a \emph{subgradient} of $\sigma$ at each point $y \in
\mathbb{R}^n$.  This derivation, along with a more formal discussion
of the non-differentiability of $\sigma$, can be found in the
Appendix.

The theory of non-differentiable, convex optimisation is perhaps less
well-known than its differentiable counterpart, but a fundamental
contribution was made by~\citet{Shor1985} with his introduction of the
subgradient method for minimising non-differentiable, convex functions
defined on Euclidean spaces.  A slightly specialised version of his
Theorem~2.2 gives that if $\partial \sigma(y)$ is a subgradient of
$\sigma$ at $y$, then for any $y^{(0)} \in \mathbb{R}^n$, the sequence
generated by the formula
\[ y^{(\ell+1)} = y^{(\ell)} - h_{\ell+1}\frac{\partial \sigma
(y^{(\ell)})}{\|\partial \sigma (y^{(\ell)})\|}
\] has the property that either there exists an index $\ell^*$ such
that $y^{(\ell^*)} = y^*$, or $y^{(\ell)} \rightarrow y^*$ and
$\sigma(y^{(\ell)}) \rightarrow \sigma(y^*)$ as $\ell \rightarrow
\infty$, provided we choose the step lengths $h_{\ell}$ so that $h_\ell \rightarrow 0$
as $\ell \rightarrow \infty$, but $\sum_{\ell = 1}^\infty h_\ell =
\infty$.

Shor recognised, however, that the convergence of this algorithm could
be slow in practice, and that although appropriate step size selection
could improve matters somewhat, the convergence would never be better
than linear (compared with quadratic convergence for Newton's method
near the optimum -- see \citet[][Section~9.5]{BoydVandenberghe2004}).
Slow convergence can be caused by taking at each stage a step in a
direction nearly orthogonal to the direction towards the optimum,
which means that simply adjusting the step size selection scheme will
never produce the desired improvements in convergence rate.

One solution \citep[][Chapter~3]{Shor1985} is to attempt to shrink the
angle between the subgradient and the direction towards the minimum
through a (necessarily nonorthogonal) linear transformation, and
perform the subgradient step in the transformed space.  By analogy
with Newton's method for smooth functions, an appropriate
transformation would be an approximation to the inverse of the Hessian
matrix at the optimum.  This is not possible for nonsmooth problems,
because the inverse might not even exist (and will not exist at points
at which the function is not differentiable, which may include the
optimum).

Instead, we perform a sequence of dilations in the direction of the
difference between two successive subgradients, in the hope of
improving convergence in the worst-case scenario of steps nearly
perpendicular to the direction towards the minimiser. This variant,
which has become known as Shor's $r$-algorithm, has been implemented
in \citet{KappelKuntsevich2000}.  Accompanying software
\texttt{SolvOpt} is available from
\url{http://www.uni-graz.at/imawww/kuntsevich/solvopt/}.

Although the formal convergence of the $r$-algorithm has not been
proved, we agree with the authors' claims that it is robust, efficient
and accurate.  Of course, it is clear that if we terminate the
$r$-algorithm after any finite number of steps and apply the original
Shor algorithm using our terminating value of $y$ as the new starting
value, then formal convergence is guaranteed.  We have not found it 
necessary to run the original Shor algorithm after termination of the 
$r$-algorithm in practice.

If $(y^{(\ell)})$ denotes the sequence of vectors in $\mathbb{R}^n$
produced by the $r$-algorithm, we terminate when 
\begin{itemize}
\item $|\sigma(y^{(\ell+1)}) - \sigma(y^{(\ell)})| \leq \delta$ 
\item  $|y_i^{(\ell+1)} - y_i^{(\ell)}| \leq \epsilon$ for $i = 1, \ldots, n$
\item  $|1 - \int \exp\{\bar{h}_{y^{(\ell)}}(x)\} \, dx| \leq \eta$ 
\end{itemize}
for some small $\delta, \epsilon \textrm{ and } \eta >0$.  The first
two termination criteria follow \citet{KappelKuntsevich2000}, while the
third is based on our knowledge that the true optimum corresponds to a
density (Section
\ref{Sec:ExistUnique}).  As default values, and throughout this paper,
we took $\delta = 10^{-8}$ and $\epsilon = \eta = 10^{-4}$.

Table~\ref{Table:Times} gives approximate running times and number of
iterations of Shor's $r$-algorithm required for different sample sizes
and dimensions on an ordinary desktop computer (1.8GHz, 2GB RAM).
Unsurprisingly, the running time increases relatively quickly with the
sample size, while the number of iterations increases approximately
linearly with $n$.  Each iteration takes longer as the dimension
increases, though it is interesting to note that the number of
iterations required for the algorithm to terminate decreases as the
dimension increases.  When $d=1$, we recommend the Active Set
algorithm of \citet{DHR2007}, which is implemented in the \texttt{R}
package \texttt{logcondens} \citep{RufibachDumbgen2006}.

\begin{table}
  \caption{\label{Table:Times} Approximate running times (with number of iterations in brackets)  
 for computing the maximum likelihood estimator of a log-concave density}
    \centering
\begin{tabular}{lllll}
      & $n=100$       & $n=500$ & $n=1000$ & $n=2000$ \\
      $d=2$ & 1.5 secs (260) & 50 secs (1270) & 4 mins (2540) & 24 mins (5370) \\ 
      $d=3$ & 6 secs (170) & 100 secs (820) & 7 mins (1530) & 44 mins (2740) \\
      $d=4$ & 23 secs (135) & 670 secs (600) & 37 mins (1100) & 224 mins (2060) 
\end{tabular}
\end{table}

\section{Finite sample performance}
\label{Sec:Simulations}

%
%
%
\begin{table}
  \caption{\label{mise2d}Mean integrated squared error estimates 
   (with standard errors in brackets where applicable; $d = 2$)} 
\centering
  \subtable[Independent Normal]{
    \begin{tabular}{llll}
      $n$ & \texttt{LogConcDEAD}   & \textbf{Kernel (opt MISE)} & \textbf{Kernel (LSCV)} \\
      100  &0.00620(0.000222)   & 0.00431 & 0.00622(0.000383) \\
      500  &0.00161(0.0000514)  & 0.00164 & 0.00199(0.0000844) \\
      1000 &0.000983(0.0000289) & 0.00106 & 0.00122(0.0000495) \\
      2000 &0.000599(0.0000155) & 0.000686& 0.000803(0.0000276) \\
    \end{tabular}
  } \\

  \subtable[Dependent Normal]{ 
    \begin{tabular}{llll}
      $n$  & \texttt{LogConcDEAD}  & \textbf{Kernel (opt MISE)}
 & \textbf{Kernel (LSCV)} \\
      100  & 0.00607(0.000283)  & 0.00440 &  0.00827(0.000583) \\
      500  & 0.00168(0.0000573) & 0.00167 &  0.00240(0.000122) \\
      1000 & 0.00100(0.0000295) & 0.00108 &  0.00142(0.0000662) \\
      2000 & 0.000608(0.0000154)& 0.000700&  0.000868(0.0000331) \\
\end{tabular}
} \\
\subtable[$\Gamma(2,1)$ (independent components)]{
  \begin{tabular}{llll}
    $n$  & \texttt{LogConcDEAD}   & \textbf{Kernel (opt AMISE)} 
    & \textbf{Kernel (LSCV)} \\
    100  & 0.00588(0.000222)   & 0.00644 & 0.00800(0.000339) \\
    500  & 0.00143(0.0000478)  & 0.00220 & 0.00291(0.0000687) \\
    1000 & 0.000802(0.0000236) & 0.00139 & 0.00194(0.0000456) \\
    2000 & 0.000451(0.0000110) & 0.000874& 0.00130(0.0000209) \\
  \end{tabular}
} \\

\subtable[Normal location mixture, $\|\mu\| = 1$]{
  \begin{tabular}{llll}
    $n$  & \texttt{LogConcDEAD}   & \textbf{Kernel (opt MISE)} 
    & \textbf{Kernel (LSCV)} \\

    100  &0.00504(0.000206)    & 0.00384 & 0.00515(0.000195)  \\
    500  &0.00136(0.0000745)    & 0.00145 &  0.00179(0.0000515) \\
    1000 &0.000747(0.0000622)   & 0.000945 &  0.00116(0.0000376) \\
    2000 &0.000543(0.0000553)    & 0.000610 & 0.000683(0.0000121) \\

  \end{tabular}
} \\
\subtable[Normal location mixture, $\|\mu\| = 2$]{
  \begin{tabular}{llll}
    $n$  & \texttt{LogConcDEAD}   & \textbf{Kernel (opt MISE)} 
    & \textbf{Kernel (LSCV)} \\
    100  & 0.00434(0.00158)   & 0.00304 &  0.00514(0.000322) \\
    500  & 0.000996(0.0000622)   & 0.00117 &  0.00146(0.000442) \\
    1000 & 0.000640(0.0000502)   & 0.000760 & 0.000880(0.000176)  \\
    2000 & 0.000445(0.0000455)    & 0.000492 & 0.000583(0.0000192) \\
  \end{tabular}
} \\

\subtable[Normal location mixture, $\|\mu\| = 3$]{
 \begin{tabular}{llll}
    $n$  & \texttt{LogConcDEAD}   & \textbf{Kernel (opt MISE)} 
    & \textbf{Kernel (LSCV)} \\

    100  &  0.00467(0.000139)   & 0.00326 &  0.00484(0.000244) \\
    500  &  0.00173(0.0000522)  & 0.00126 &  0.00150(0.000363) \\
    1000 &  0.00122(0.0000456)  & 0.000819 &  0.000925(0.0000131) \\
    2000 &  0.00105(0.0000340)  & 0.000530 &  0.000577(0.0000651) \\
  \end{tabular}
} 

\end{table}

\begin{table}
  \caption{\label{mise3d}Mean integrated squared error estimates 
   (with standard errors in brackets where applicable; $d = 3$)} 

 \centering
 
\subtable[Independent Normal]{
  \begin{tabular}{llll}
    $n$  & \texttt{LogConcDEAD}   & \textbf{Kernel (opt MISE)} 
    & \textbf{Kernel (LSCV)} \\
    100  & 0.00426(0.000131)  & 0.00240  &  0.00505(0.000279) \\
    500  & 0.000835(0.0000302) & 0.00106  &  0.00143(0.0000338)\\
    1000 & 0.000442(0.0000236) & 0.000737 &  0.000888(0.0000139)\\
    2000 & 0.000304(0.0000238)& 0.000508 &  0.000579(0.00000985)\\
  \end{tabular}
} \\

\subtable[Dependent Normal]{
  \begin{tabular}{llll}
    $n$  & \texttt{LogConcDEAD}   & \textbf{Kernel (opt MISE)} 
    & \textbf{Kernel (LSCV)} \\
    100  &  0.00467(0.000147)  & 0.00254  & 0.00550(0.000361) \\
    500  &  0.000812(0.0000301)  & 0.00112  & 0.00152(0.0000367) \\
    1000 &  0.000431(0.0000249) & 0.000778 &  0.000922(0.0000145)\\
    2000 & 0.000304(0.0000233)   & 0.000537 & 0.000603(0.00000676) \\
  \end{tabular}
} \\

\subtable[$\Gamma(2,1)$ (independent components)]{
 \begin{tabular}{llll}
    $n$  & \texttt{LogConcDEAD}   & \textbf{Kernel (opt AMISE)} 
    & \textbf{Kernel (LSCV)} \\

    100  & 0.00365(0.000142)   & 0.00344  & 0.0741(0.00400) \\
    500  & 0.000779(0.0000243) & 0.00136  &  0.00192(0.0000518) \\
    1000 & 0.000538(0.000104)  & 0.000922 &0.00123(0.0000262)  \\
    2000 & 0.000292(0.0000414) & 0.000622 &0.000849(0.0000228) \\

  \end{tabular}
} \\

\subtable[Normal location mixture, $\|\mu\| = 1$]{
  \begin{tabular}{llll}
    $n$  & \texttt{LogConcDEAD}   & \textbf{Kernel (opt MISE)} 
    & \textbf{Kernel (LSCV)} \\
    100  & 0.00395(0.000124)   & 0.00214 & 0.00446(0.000242)   \\
    500  & 0.000743(0.0000272) & 0.000946 & 0.00124(0.0000298) \\
    1000 & 0.000446(0.0000218) & 0.000656 & 0.000822(0.0000179)  \\
    2000 & 0.000265(0.0000202) & 0.000452 & 0.000508(0.00000537) \\
  \end{tabular}
} \\
 
\subtable[Normal location mixture, $\|\mu\| = 2$]{
  \begin{tabular}{llll}
    $n$  & \texttt{LogConcDEAD}   & \textbf{Kernel (opt MISE)} 
    & \textbf{Kernel (LSCV)} \\
    100  & 0.00319(0.000100)     & 0.00168 &0.00371(0.000203)  \\
    500  & 0.000596(0.0000231)   & 0.000748 & 0.00103(0.0000340)  \\
    1000 &  0.000329(0.0000173)  & 0.000520 &  0.000656(0.0000160) \\
    2000 &  0.000220(0.0000171)  & 0.000358 & 0.000410(0.00000519)  \\
  \end{tabular}
} \\

\subtable[Normal location mixture, $\|\mu\| = 3$]{
 \begin{tabular}{llll}
    $n$  & \texttt{LogConcDEAD}   & \textbf{Kernel (opt MISE)} 
    & \textbf{Kernel (LSCV)} \\
    100  &  0.00328(0.0000930)   & 0.00166 &  0.00296(0.000120)  \\
    500  &  0.000803(0.0000184)  & 0.000751 & 0.000998(0.000254) \\
    1000 &  0.000552(0.0000169)  & 0.000525 & 0.000613(0.0000892)  \\
    2000 &  0.000401(0.0000133)  & 0.000364 & 0.000404(0.00000488) \\
  \end{tabular}
} 
\end{table}

Our simulation study considered, for $d = 2$ and $3$, the following densities:
\begin{enumerate}
\item[(a)] standard normal, $\phi_d \equiv \phi_{d,I}$
\item[(b)] dependent normal, $\phi_{d,\Sigma}$, with $\Sigma_{ij} = \mathbbm{1}_{\{i=j\}} + 0.2\mathbbm{1}_{\{i \neq j\}}$
\item[(c)] the joint density of independent $\Gamma(2,1)$ components  
\item[(d-f)] the normal location mixture $0.6 \phi_d(\cdot) + 0.4 \phi_d(\cdot - \mu)$ for (d) $\|\mu\| = 1$, (e) $\|\mu\| = 2$, (f) $\|\mu\| = 3$.  An application of Proposition~\ref{Prop:LCProperties} gives that such 
a normal location mixture is log-concave if and only if $\|\mu\| \leq 2$.
\end{enumerate}
In Tables~\ref{mise2d} and \ref{mise3d} we present, for each density
and for four different sample sizes, an estimate of the mean
integrated squared error (MISE) of the nonparametric maximum
likelihood estimator based on 100 Monte Carlo iterations.  We also
show the MISE for the kernel density estimates with a Gaussian kernel
and, for all of the normal and mixture of normal examples, the choice
of bandwidth that minimises the MISE.  In the gamma example, exact
MISE calculations are not possible, so we took the bandwidth that
minimises the asymptotic mean integrated squared error (AMISE).  These
optimal bandwidths can be computed using the formulae in
\citet[Sections~4.3 and~4.4]{WandJones1995}.  As minimisation of the
expressions for both the MISE and the AMISE requires knowledge of
certain functionals of the true density that would be unknown in
practice, we also provide a comparison with an empirical bandwidth
selector based on least squares cross validation (LSCV)
\citep[Section~4.7]{WandJones1995}.  The LSCV bandwidths were computed
using the~\texttt{ks} package
\citep{Duong2007} in~\texttt{R}, and we used the option of
constraining the bandwidth matrices to be diagonal in cases~(a)
and~(c) where the components are independent.

We see that in cases (a)-(e) the log-concave maximum likelihood
estimator has a smaller MISE than the kernel estimate with bandwidth
chosen by LSCV, and at least for moderate and large sample sizes, the
difference is quite dramatic.  Even more remarkably, in these cases
the log-concave estimator also outperforms the kernel estimate with
optimally chosen bandwidth when the sample size is not too small.  It
seems that for small sample sizes, the fact that the convex hull of
the data is rather small hinders the performance of the log-concave
estimator, but that this effect is reduced as the sample size
increases.  The log-concave estimator copes well with the dependence
in case~(b), and it also deals particularly impressively with
case~(c), where the true density decays to zero at the boundary of the
positive orthant.

In case~(f), where the log-concavity assumption is violated, the
performance of our estimator is not as good as the kernel estimate
with the optimally chosen bandwidth, but is still comparable in most
cases with the LSCV method.  One would not expect the MISE of
$\hat{f}_n$ to approach zero as $n \rightarrow \infty$ if
log-concavity is violated, and in fact we conjecture that in this case
the log-concave maximum likelihood estimator will converge to the
density $f^*$ that minimises the Kullback--Leibler divergence $d(f_0
\,
\| \, f) = \int f_0(x) \log \frac{f_0(x)}{f(x)} \, dx$ over $f \in
\mathcal{F}_0$.  Such a result would be interesting for robustness purposes, 
because it could be interpreted as saying that provided the underlying
density does not violate the log-concavity assumption too seriously,
the log-concave maximum likelihood estimator is still sensible.

\section{Clustering example}
\label{Sec:Clustering}

In a recent paper, \citet{ChangWalther2008} introduced an algorithm
which combines the univariate log-concave maximum likelihood estimator
with the EM algorithm \citep{DLR1977}, to fit a finite mixture density
of the form
\begin{equation}
\label{Eq:mixture}
f(x) = \sum_{j=1}^p \pi_j f_j(x),
\end{equation}
where the mixture proportions $\pi_1,\ldots,\pi_p$ are positive and
sum to one, and the component densities $f_1,\ldots,f_p$ are
univariate and log-concave.  The method is an extension of the
standard Gaussian EM algorithm, e.g. \citet{FraleyRaftery2002}, which
assumes that each component density is normal.  Once estimates
$\hat{\pi}_1,\ldots,\hat{\pi}_p,\hat{f}_1,\ldots,\hat{f}_p$ have been
obtained, clustering can be carried out by assigning to the $j$th
cluster those observations $X_i$ for which $j = \mathrm{argmax}_r \,
\hat{\pi}_r \hat{f}_r(X_i)$.  \citet{ChangWalther2008} show
empirically that in cases where the true component densities are
log-concave but not normal, their algorithm tends to make considerably
fewer misclassifications and have smaller mean absolute error in the
mixture proportion estimates than the Gaussian EM algorithm, with very
similar performance in cases where the true component densities are
normal.

Owing to the previous lack of an algorithm for computing the maximum
likelihood estimator of a multidimensional log-concave density,
\citet{ChangWalther2008} discuss an extension of the model
in~(\ref{Eq:mixture}) to a multivariate context where the univariate
marginal densities of each component in the mixture are assumed to be
log-concave, and the dependence structure within each component
density is modelled with a normal copula.  Now that we are able to
compute the maximum likelihood estimator of a multidimensional
log-concave density, we can carry this method through to its natural
conclusion.  That is, in the finite mixture model~(\ref{Eq:mixture})
for a multidimensional log-concave density $f$, we simply assume that
each of the component densities $f_1,\ldots,f_p$ is log-concave. An
interesting problem that we do not address here that of finding
appropriate conditions under which this model is identifiable --
see~\citet[Section~3.1]{TSM1985} for a nice discussion.

\subsection{EM algorithm}
An introduction to the EM algorithm can be found in
\citet{McLachlanKrishnan1997}.  Briefly, given current estimates of
the mixture proportions and component densities
$\hat{\pi}_1^{(\ell)},\ldots,\hat{\pi}_p^{(\ell)},
\hat{f}_1^{(\ell)},\ldots,\hat{f}_p^{(\ell)}$
at the $\ell$th iteration of the algorithm, we update the estimates of
the mixture proportions by setting $\hat{\pi}_j^{(\ell+1)} =
n^{-1}\sum_{i=1}^n \hat{\theta}_{i,j}^{(\ell)}$ for $j=1,\ldots,p$,
where
\[
\hat{\theta}_{i,j}^{(\ell)} =
\frac{\hat{\pi}_j^{(\ell)}\hat{f}_j^{(\ell)}(X_i)}
{\sum_{r=1}^p \hat{\pi}_r^{(\ell)}\hat{f}_r^{(\ell)}(X_i)}
\]
is the current estimate of the posterior probability that the $i$th
observation belongs to the $j$th component.  We then update the
estimates of the component densities in turn using the algorithm
described in Section~\ref{Sec:Computation}, choosing
$\hat{f}_j^{(\ell+1)}$ to be the log-concave density $f_j$ that
maximises
\[
\sum_{i=1}^n \hat{\theta}_{i,j}^{(\ell)} \log f_j(X_i).
\]
The incorporation of the weights
$\hat{\theta}_{1,j}^{(\ell)},\ldots,\hat{\theta}_{n,j}^{(\ell)}$ in
the maximisation process presents no additional complication, as is
easily seen by inspecting the proof of Theorem~\ref{Thm:ExistUnique}.
As usual with methods based on the EM algorithm, although the
likelihood increases at each iteration, there is no guarantee that the
sequence converges to a global maximum.  In fact, it can happen that
the algorithm produces a sequence that approaches a degenerate
solution, corresponding to a component concentrated on a single
observation, so that the likelihood becomes arbitrarily high.  The
same issue can arise when fitting mixtures of Gaussian densities, and
in this context \citet{FraleyRaftery2002} suggest that a Bayesian
approach can alleviate the problem in these instances by effectively
smoothing the likelihood.  In general, it is standard practice to
restart the algorithm from different initial values, taking the
solution with the highest likelihood.

\subsection{Breast cancer example}
We illustrate the log-concave EM algorithm on the Wisconsin breast
cancer data set of \citet{SWM1993}, available on the UCI Machine
Learning Repository website \citep{AsuncionNewman2007}:

\url{http://archive.ics.uci.edu/ml/datasets/Breast+Cancer+Wisconsin+
\%28Diagnostic\%29}.

\begin{figure}
\centering 
\psfrag{Malignant}{\footnotesize{Malignant}}
\psfrag{Benign}{\footnotesize{Benign}} \psfrag{0.5}{\scriptsize{0.5}}
\psfrag{1.0}{\scriptsize{1.0}} \psfrag{1.5}{\scriptsize{1.5}}
\psfrag{2.0}{\scriptsize{2.0}} \psfrag{2.5}{\scriptsize{2.5}}
\psfrag{1}{\scriptsize{1}} \psfrag{2}{\scriptsize{2}}
\psfrag{3}{\scriptsize{3}} \psfrag{4}{\scriptsize{4}}
\psfrag{5}{\scriptsize{5}} \psfrag{X}{\vspace{1pt}\footnotesize{X}}
\psfrag{Y}{\footnotesize{Y}} \psfrag{Correct}{\footnotesize{Correct}}
\psfrag{Incorrect}{\footnotesize{Incorrect}} 
\mbox{ 
\subfigure[Data]{\includegraphics[scale=0.25,angle=-90]{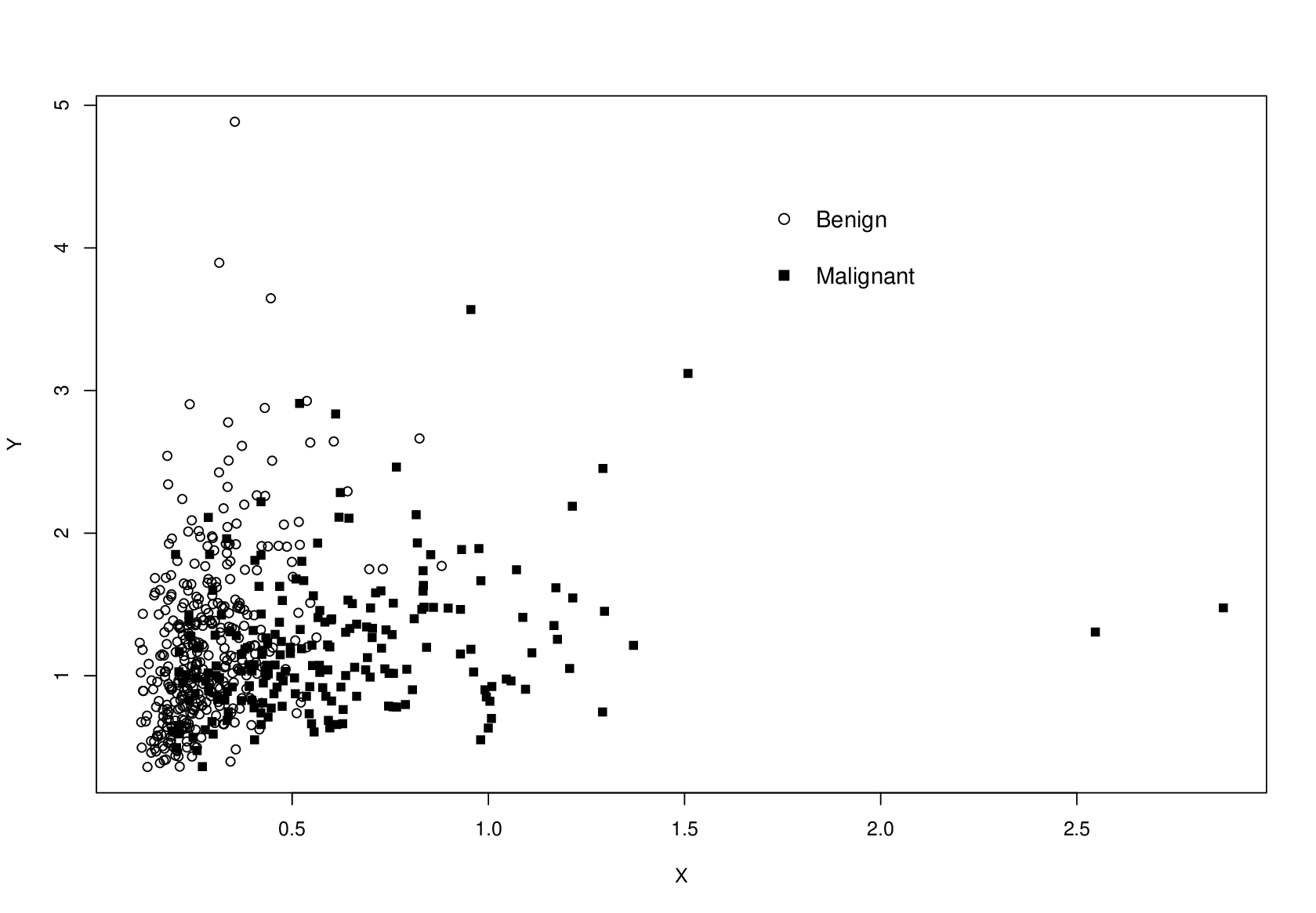} } 
\subfigure[Gaussian
mixture classification]{
\includegraphics[scale=0.25,angle=-90]{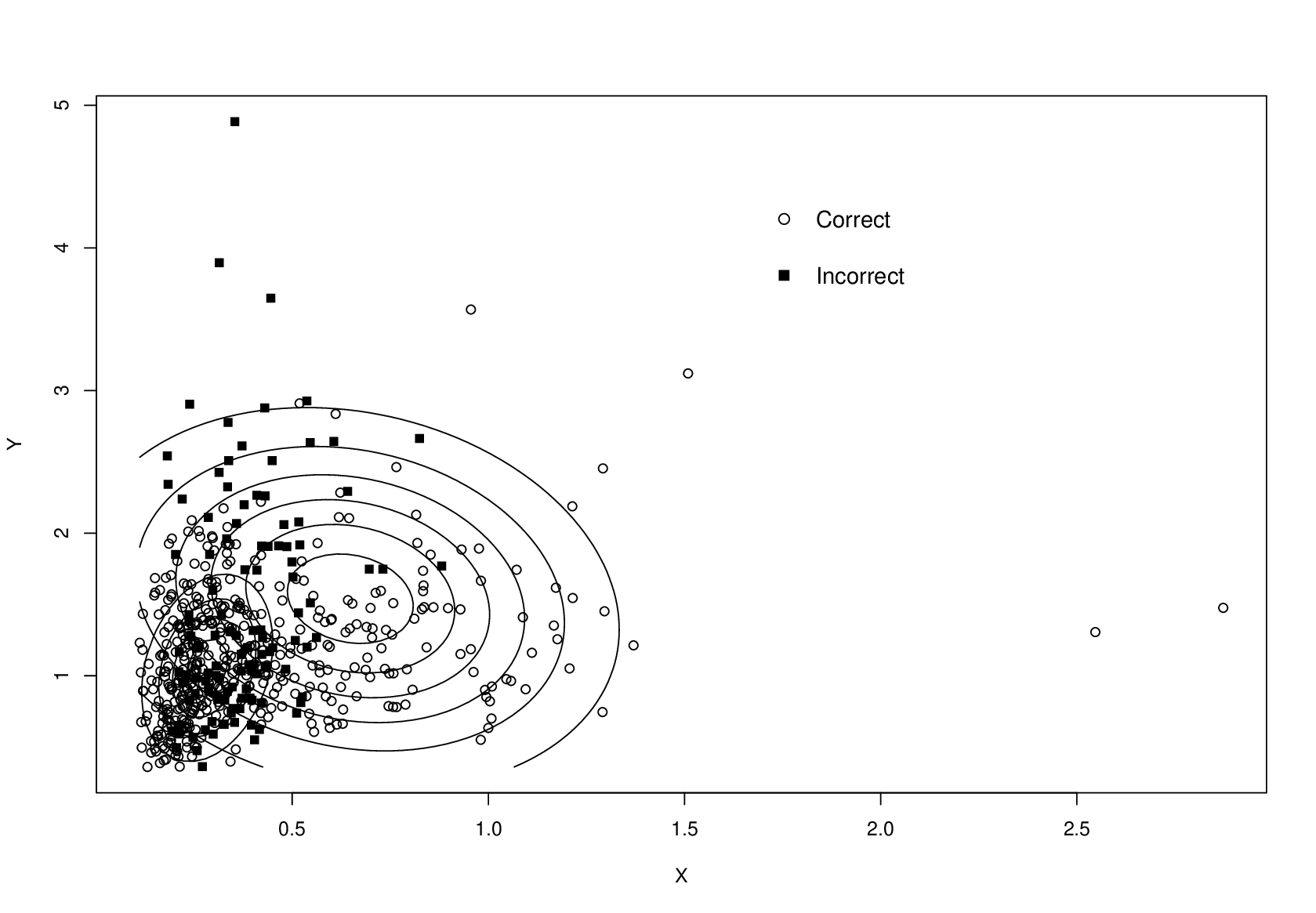} } } 
\mbox{
\subfigure[Log-concave mixture classification]{
\includegraphics[scale=0.25,angle=-90]{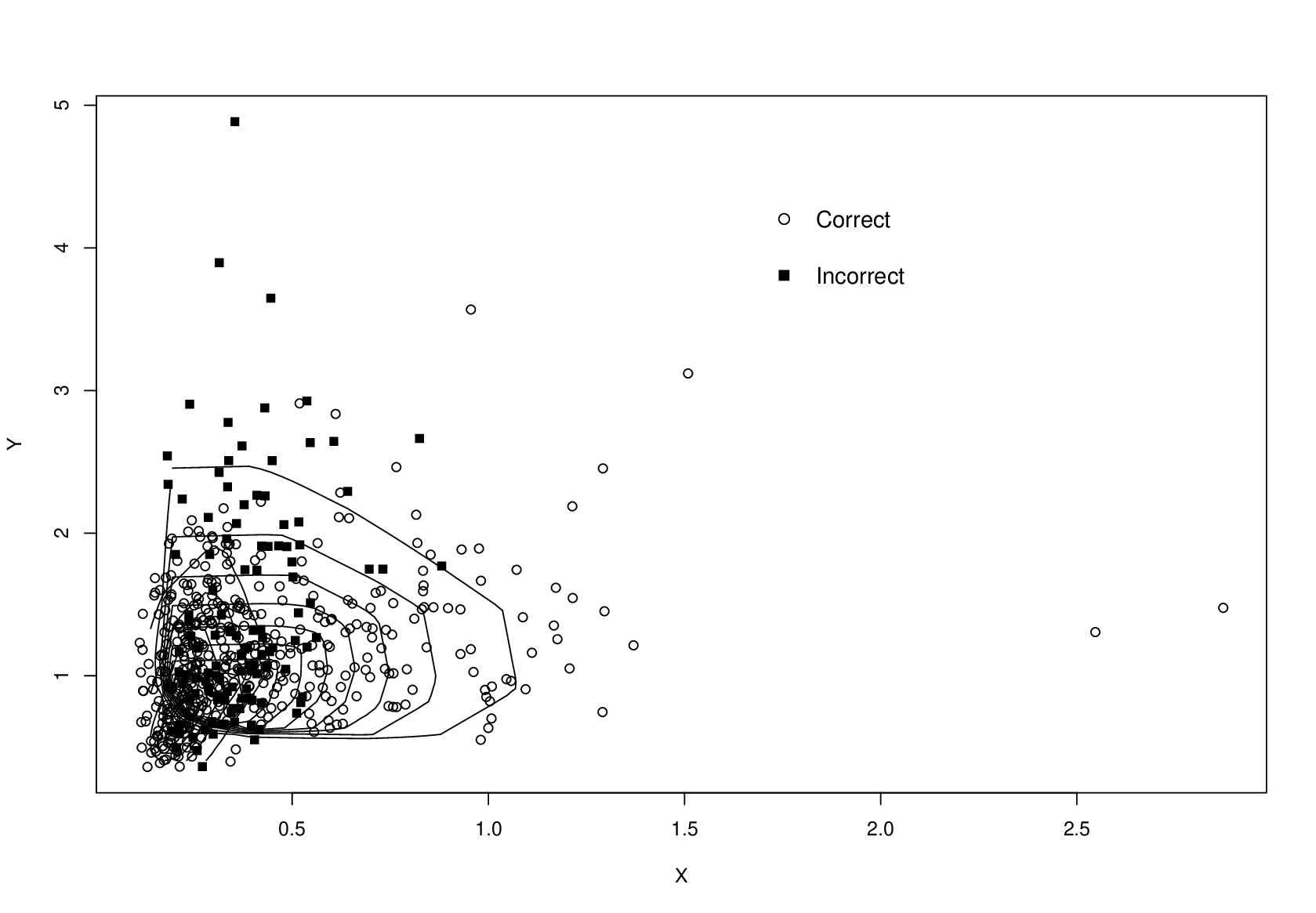} } \subfigure[Estimated
log-concave mixture]{ \vspace{5cm}
\includegraphics[scale=0.46,angle=-90]{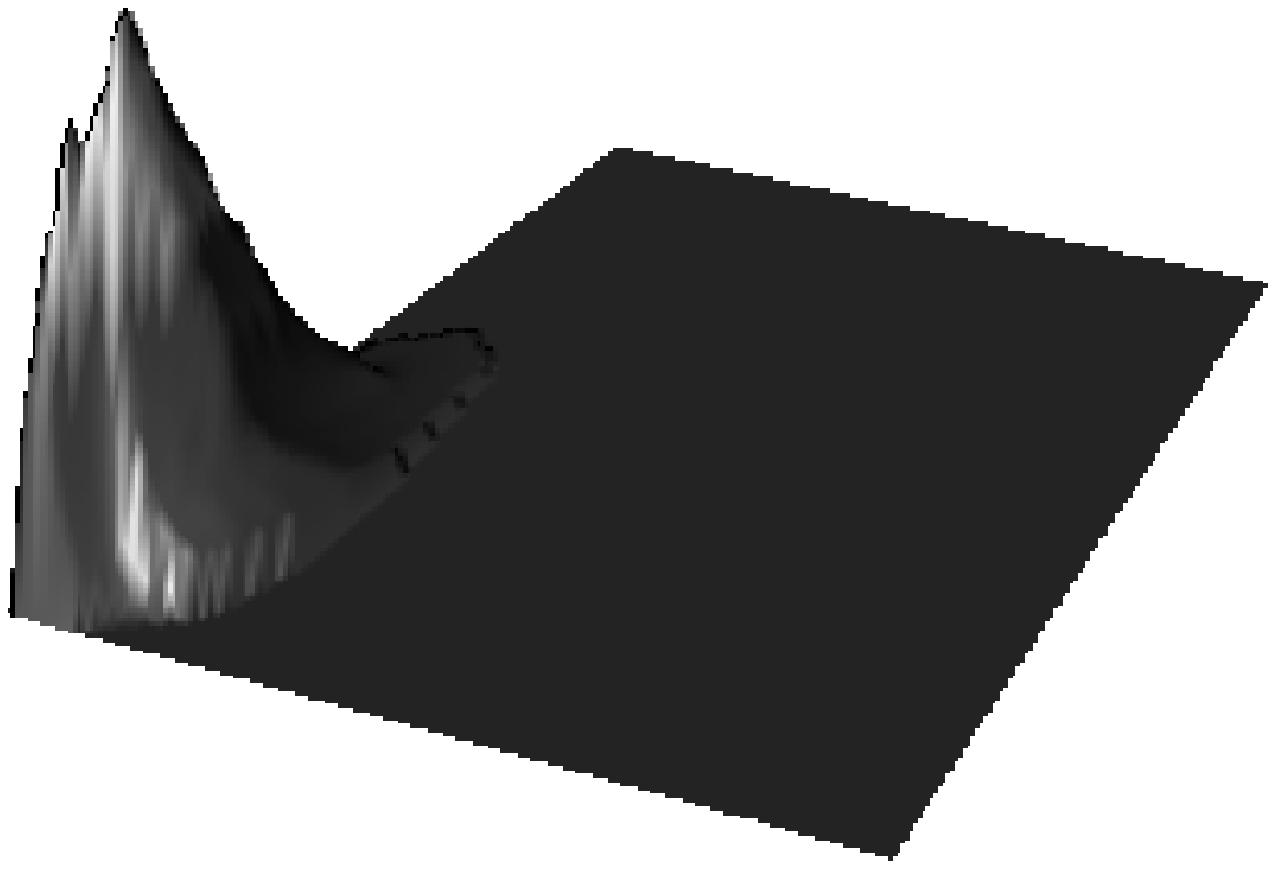} } }
\caption{\label{Fig:BreastCancer}Panel~(a) plots the Wisconsin breast
cancer data, with benign cases as solid squares and malignant ones as
open circles.  Panel~(b) gives a contour plot together with the
misclassified instances from the Gaussian EM algorithm, while the
corresponding plot obtained from the log-concave EM algorithm is given
in Panel~(c).  Panel~(d) plots the fitted mixture distribution from
the log-concave EM algorithm.}
\end{figure}

The data set was created by taking measurements from a digitised image
of a fine needle aspirate of a breast mass, for each of 569
individuals, with 357 benign and 212 malignant instances.  We study
the problem of trying to diagnose (cluster) the individuals based on
the standard errors of two of the measurements, namely the radius of
the cell nucleus (mean of distances from center to points on the
perimeter, $X$) and its texture (standard deviation of grey-scale
values, $Y$).  The data are presented in
Figure~\ref{Fig:BreastCancer}(a).  In fact, the full data set consists
of 30 measurements for each patient, representing the mean, standard
error and `worst' (mean of the three largest values) of 10 different
features computed for each cell nucleus in the image.  Since one would
reasonably expect the means of each feature to be approximately
normally distributed, and hence the Gaussian EM algorithm to be
appropriate, we took the standard errors of the first two measurements
to illustrate the log-concave EM algorithm methodology.

It is important also to note that although for this particular data
set we do know whether a particular instance is benign or malignant,
we did not use this information in fitting our mixture model.  Instead
this information was only used afterwards to assess the performance of
the method, as reported below.  Thus we are studying a clustering (or
unsupervised learning) problem, by taking a classification (or
supervised learning) data set and `covering up the labels' until it
comes to performance assessment.

The skewness in the data suggests that the mixture of Gaussians model
may be inadequate, and in Figure~\ref{Fig:BreastCancer}(b) we show the
contour plot and misclassified instances from this model.  The
corresponding plot obtained from the log-concave EM algorithm is given
in Figure~\ref{Fig:BreastCancer}(c), while
Figure~\ref{Fig:BreastCancer}(d) plots the fitted mixture distribution
from the log-concave EM algorithm.  For this example, the number of
misclassified instances is reduced from 144 with the Gaussian EM
algorithm to 121 with the log-concave EM algorithm.

In some examples, it will be necessary to estimate $p$, the number of
mixture components.  In the general context of model-based clustering,
\citet{FraleyRaftery2002} cite several possible approaches for this
purpose, including methods based on resampling
\citep{McLachlanBasford1988} and an information criterion
\citep{Bozdogan1994}.  Further research will be needed to ascertain
which of these methods is most appropriate in the context of
log-concave component densities.

\section{Plug-in estimation of functionals, sampling and the
bootstrap}
\label{Sec:Sample}

Suppose $X$ has density $f$.  Often, we are less interested in
estimating a density directly than in estimating some functional
$\theta(f)$.  Examples of functionals of interest (some of which were
given in Section~\ref{Sec:Intro}), include:
\begin{enumerate}
\item[(a)] $\mathbb{P}(\|X\| \geq 1) = \int f(x)\mathbbm{1}_{\{\|x\| \geq
1\}} \, dx$
\item[(b)] Moments, such as $\mathbb{E}(X) = \int x f(x) \, dx$, or 
$\mathbb{E}(\|X\|^2) = \int \|x\|^2 f(x) \, dx$
\item[(c)] The differential entropy of $X$ (or $f$), defined by $H(f) =
-\int f(x) \, \log f(x) \, dx$
\item[(d)] The $100(1-\alpha)\%$ highest density region, defined by
$R_{\alpha} = \{x \in \mathbb{R}^d: f(x) \geq f_\alpha\}$, where
$f_{\alpha}$ is the largest constant such that $\mathbb{P}(X \in
R_{\alpha}) \geq 1 - \alpha$.  \citet{Hyndman1996} argues that this
is an informative summary of a density; note that subject to a minor
restriction on $f$, we have $\int f(x)\mathbbm{1}_{\{f(x) \geq
f_{\alpha}\}} \, dx = 1 - \alpha$.
\end{enumerate} 
Each of these may be estimated by the corresponding functional
$\theta(\hat{f}_n)$ of the log-concave maximum likelihood estimator.
In examples~(a) and~(b) above, $\theta(f)$ may also be written as a
functional of the corresponding distribution function $F$,
e.g. $\mathbb{P}(\|X\| \geq 1) = \int \mathbbm{1}_{\{\|x\| \geq 1\}}
dF(x)$.  In such cases, it is more natural to use the plug-in
estimator based on the empirical distribution function, $\hat{F}_n$,
of the sample $X_1,\ldots,X_n$, and indeed in our simulations we found
that the log-concave plug-in estimator did not offer an improvement 
on this method.  In the other examples, however, an empirical distribution 
function plug-in estimator is not available, and the log-concave plug-in 
estimator is a potentially attractive procedure. 

\subsection{Monte Carlo estimation of functionals and sampling from the density estimate}
\label{Sec:MC}
For some functionals we can compute $\hat{\theta} = \theta(\hat{f}_n)$
analytically.  If this is not possible, but we can write $\theta(f) =
\int f(x) g(x) \, dx$, we may approximate $\hat{\theta}$ by
\[\hat{\theta}_B = \frac{1}{B} \sum_{b=1}^B g(X_b^*),\]
for some (large) $B$, where $X_1^*, \ldots, X^*_B$ are independent
samples from $\hat{f}_n$.  Conditional on $X_1,\ldots,X_n$, the strong
law of large numbers gives that $\hat{\theta}_B
\stackrel{a.s.}{\rightarrow} \hat{\theta}$ as $B
\rightarrow \infty$.  In practice, even when analytic calculation of
$\hat{\theta}$ was possible, this method was found to be fast and
accurate.

In order to use this Monte Carlo procedure, we must be able to sample
from $\hat{f}_n$.  Fortunately, this can be done efficiently using the 
following rejection sampling procedure.  As in Section
\ref{Sec:Computation}, for $j \in J$ let $A_j$ be the $d \times d$
matrix whose $l$th column is $X_{j_{l+1}} - X_{j_1}$ for
$l=1,\ldots,d$, and let $\alpha_j = X_{j_1}$, so that $w \mapsto A_j w
+ \alpha_j$ maps the unit simplex $T_d$ to $C_{n,j}$.  Recall
that $\log \hat{f}_n(X_i) = y_i^*$, and let $z_j = (z_{j,1},\ldots,z_{j,d})$, 
where $z_{j,l} = y_{j_{l+1}}^* - y_{j_1}^*$ for $l=1,\ldots,d$. Write
\[q_j = \int_{C_{n,j}} \hat{f}_n(x) \, dx.\] We may then draw an
observation $X^*$ from $\hat{f}_n$ as follows:
\begin{enumerate}
\item Select $j^* \in J$, selecting $j^*=j$ with probability $q_j$
\item Select $w \sim \textrm{Unif}(T_d)$ and $u \sim
  \textrm{Unif}([0,1])$ independently.  If \[ u < \frac{\exp({\langle
  w, z_{j^*}\rangle})}{\max_{v \in T_d}\exp({\langle v,z_{j^*}
  \rangle})},\] accept the point and set $X^* = A_j w + \alpha_j$.
  Otherwise, repeat (ii).
\end{enumerate}

\subsection{Simulation study}

In this section we illustrate some simple applications of this
technique to functionals~(c) and (d) above, using the Monte Carlo
procedure and sampling scheme described in Section \ref{Sec:MC}.
Estimates are based on random samples from a $N_2(0,I)$
distribution, and we compare the performance of the
\texttt{LogConcDEAD} estimate with that of a kernel-based 
plug-in estimate, where the bandwidth matrix was chosen using our 
knowledge of the underlying density to minimise the MISE.  

Table~\ref{Table:functional}(a) gives mean squared errors (with Monte
Carlo standard errors) of the plug-in estimates of the differential
entropy.  In Table~\ref{Table:functional}(b) we study the plug-in
estimators $\hat{R}_\alpha$ of the highest density region $R_\alpha$,
and measure the quality of the estimation procedures through
$\mathbb{E}\{\mu_f(\hat{R}_\alpha \bigtriangleup R_\alpha)\}$, where
$\mu_f(A) = \int_A f(x) \, dx$ and $\bigtriangleup$ denotes set
difference.  Highest density regions can be computed once we have
approximated the sample versions of $f_\alpha$ using the density
quantile algorithm described in \citet[][Section~3.2]{Hyndman1996}.

\begin{figure}
  \centering 
  \subfigure[\texttt{LogConcDEAD} estimate]{
    \includegraphics[angle=-90,scale=0.25]{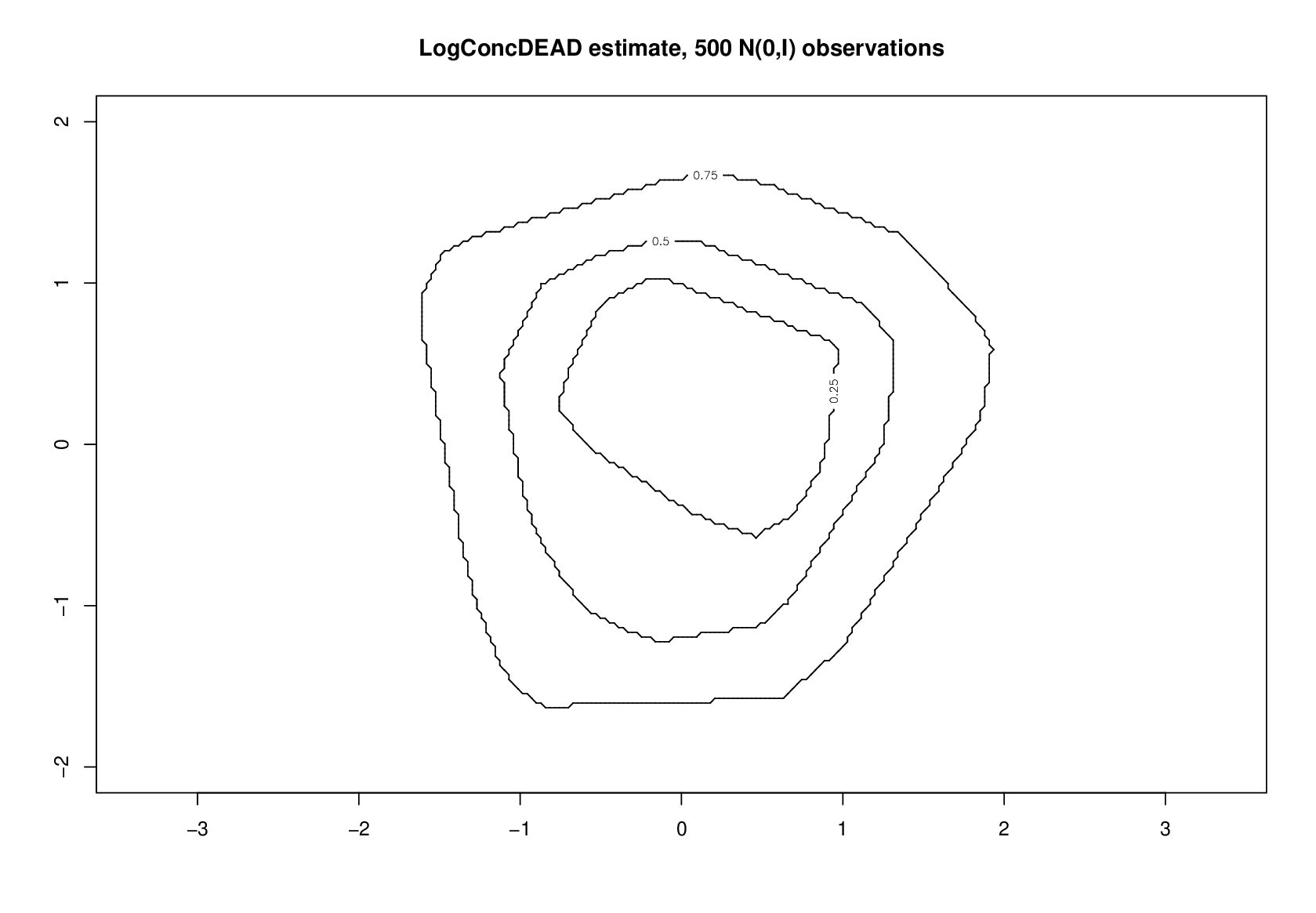}
  }
  \subfigure[True]{
    \includegraphics[angle=-90,scale=0.25]{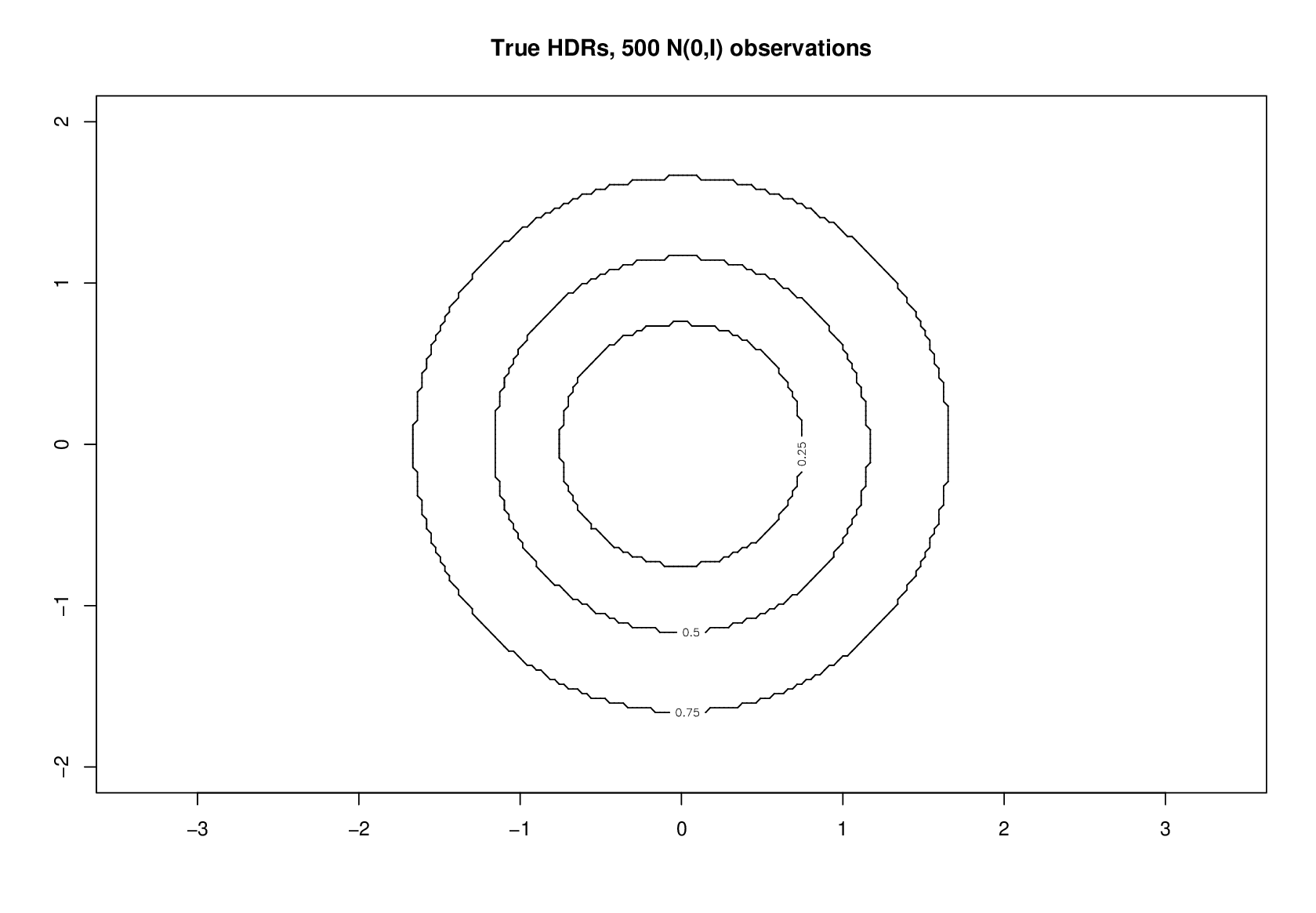}
  }
  \subfigure[Kernel estimate]{
    \includegraphics[angle=-90,scale=0.25]{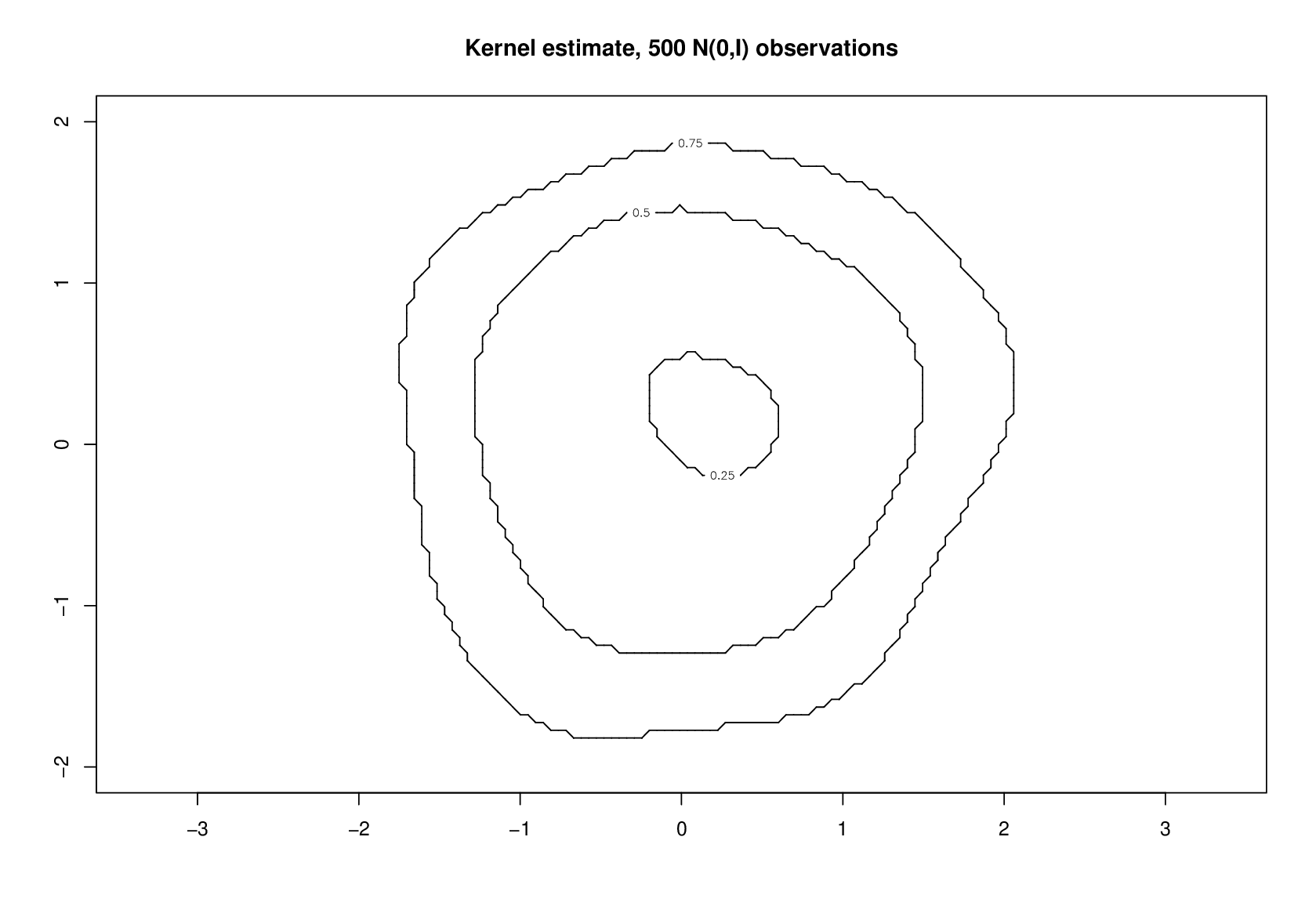}
  }
  \caption{Estimates of the 25\%, 50\% and 75\% highest density region
    from 500 observations from the $N_2(0,I)$ distribution.}
  \label{fig:HDR}
\end{figure}

For the differential entropy estimators, we find a similar pattern to
that observed in Section~\ref{Sec:Simulations}: the log-concave
plug-in estimator provides an improvement on the kernel-based
estimator for the moderate and large sample sizes in our simulations.
For the case of highest density regions, the relative performance of
the log-concave estimator is better for the estimation of smaller
density regions.  In Figure \ref{fig:HDR}, we illustrate the
estimation of three highest density regions based on 500 points from a
$N_2(0,I)$ distribution.  For comparison, a kernel-based plug-in
estimate (where the regions are not guaranteed to be convex) is also
given.

\begin{table}

\caption{\label{Table:functional}(a) gives mean squared errors for estimating 
the differential entropy of the $N_2(0,I)$ distribution; (b) gives
$\mathbb{E}\{\mu_f(\hat{R}_\alpha \bigtriangleup R_\alpha)\}$ when
estimating highest density regions.  The numbers in brackets are Monte
Carlo standard errors.}
\centering
 \subtable[Differential entropy]{
  \begin{tabular}{lll}
    $n$ & \texttt{LogConcDEAD} & Kernel \\
    100 & 0.0761(0.00629) & 0.0457(0.00304) \\
    500 & 0.00819(0.000653) & 0.0137(0.000839)\\
    1000& 0.00378(0.000391) & 0.00716(0.000581)\\
    2000& 0.00177(0.000232) & 0.00427(0.000345)\\
  \end{tabular}
 }
\\

\subtable[25\%/50\%/75\% highest density regions]{ \label{Table:hdr}
  \begin{tabular}{lll}
    $n$ & \texttt{LogConcDEAD} & Kernel \\
    100 &0.0872(0.0024)/0.110(0.0033)/0.121(0.0047)& 0.0753(0.0017)/0.0995(0.0028)/0.0959(0.0038)\\
    500 &0.0419(0.0010)/0.0587(0.0014)/0.0680(0.0022)&0.0467(0.0011)/0.0609(0.0013)/0.0637(0.0019)\\
   1000& 0.0311(0.00075)/0.0447(0.0011)/0.0536(0.0016)&0.0376(0.00095)/0.0476(0.0012)/0.0477(0.0015)\\
   2000 &0.0241(0.00054)/0.0363(0.00080)/0.0448(0.0013) &0.0322(0.00081)/0.0371(0.00098)/0.0399(0.0013) \\
  \end{tabular}


}
\end{table}

In real data examples, we are unable to assess uncertainty in our
functional estimates by taking repeated samples from the true
underlying model.  Nevertheless, the fact that we can sample from the
log-concave maximum likelihood estimator does mean that we can apply
standard bootstrap methodology to compute standard errors or
confidence intervals, for example.  Finally, we remark that the
plug-in estimation procedure, sampling algorithm and bootstrap
methodology extend in an obvious way to the case of a finite mixture
of log-concave densities.

\section{Concluding discussion}
\label{Sec:Discussion}

We have developed methodology that gives a fully automatic
nonparametric density estimate under the condition that the density is
log-concave, and shown how it may be extended to fit finite mixtures
of log-concave densities.  We have indicated a wide range of possible
applications, including classification, clustering and functional
estimation problems.  The area of shape-constrained estimation is
currently undergoing rapid growth, as evidenced by the many recent
publications cited in the penultimate paragraph of
Section~\ref{Sec:Intro}, as well as recent workshops in Oberwolfach
(November 2006), Eindhoven (October 2007) and Bristol (November 2007).
We hope that this paper will stimulate further interest and research
in the field.

As well as the continued development and refinement of the
computational algorithms and graphical displays of estimates, and
studies of theoretical performance, there remain many challenges and
interesting directions for future research.  These include:
\begin{enumerate}
\item Studying other shape constraints.  These have received some
attention for univariate data, dating back to \citet{Grenander1956},
but much less in the multivariate setting.
\item Developing both formal and informal diagnostic tools for
assessing the validity of shape constraints.
\item Assessing the uncertainty in shape-constrained nonparametric
density estimates, through confidence intervals/bands.
\item Developing analogous methodology for discrete data from
shape-constrained distributions.
\item Examining nonparametric shape constraints in regression
problems.
\item Studying methods for choosing the number of clusters in
nonparametric, shape-constrained mixture models.
\end{enumerate}

\appendix
\section{Glossary of terms and results from convex analysis and
computational geometry}
\label{Sec:Glossary}

All of the definitions and results below can be found in
\citet{Rockafellar1997} and \citet{Lee1997}.  The \emph{epigraph} of a
function $f:\mathbb{R}^k \rightarrow [-\infty,\infty)$ is the set
\[ \mathrm{epi}(f) = \{(x,\mu): x \in \mathbb{R}^k, \mu \in
\mathbb{R}, \mu \leq f(x)\}.
\] We say $f$ is \emph{concave} if its epigraph is non-empty and
convex as a subset of $\mathbb{R}^{k+1}$; note that this agrees with
the terminology of \citet{Barndorff-Nielsen1978}, but is
what~\citet{Rockafellar1997} calls a \emph{proper concave} function.
If $C$ is a convex subset of $\mathbb{R}^k$ then provided $f: C
\rightarrow [-\infty,\infty)$ is not identically $-\infty$, it is
\emph{concave} if and only if
\[ f\bigl(tx + (1-t)y\bigr) \geq t f(x) + (1-t) f(y)
\] for $x,y \in C$ and $t \in (0,1)$.  A non-negative function $f$ is
\emph{log-concave} if $\log f$ is concave, with the convention that
$\log 0 = -\infty$.  The \emph{support} of a log-concave function $f$
is $\{x \in \mathbb{R}^k: \log f(x) > -\infty\}$, a convex subset of
$\mathbb{R}^k$.

A subset $M$ of $\mathbb{R}^k$ is \emph{affine} if $tx + (1-t)y \in M$
for all $x,y \in M$ and $t \in \mathbb{R}$.  The \emph{affine hull} of
$M$, denoted $\mathrm{aff}(M)$, is the smallest affine set containing
$M$.  Every non-empty affine set $M$ in $\mathbb{R}^k$ is
\emph{parallel} to a unique subspace of $\mathbb{R}^k$, meaning that
there is a unique subspace $L$ of $\mathbb{R}^k$ such that $M = L+a$,
for some $a \in \mathbb{R}^k$.  The \emph{dimension} of $M$ is the
dimension of this subspace, and more generally the dimension of a
non-empty convex set is the dimension of its affine hull.  A finite
set of points $M = \{x_0,x_1,\ldots,x_d\}$ is \emph{affinely
independent} if $\mathrm{aff}(M)$ is $d$-dimensional.  The
\emph{relative interior} of a convex set $C$ is the interior which
results when we regard $C$ as a subset of its affine hull.  The
\emph{relative boundary} of $C$ is the set difference between its
closure and its relative interior.  If $M$ is an affine set in
$\mathbb{R}^k$, then an \emph{affine transformation} (or \emph{afffine
function}) is a function $T:M \rightarrow \mathbb{R}^k$ such that
$T\bigl(tx+(1-t)y\bigr) = tT(x) + (1-t)T(y)$ for all $x,y \in M$ and
$t \in \mathbb{R}$.

The \emph{closure} of a concave function $g$ on $\mathbb{R}^d$,
denoted $\mathrm{cl}(g)$, is the function whose epigraph is the
closure in $\mathbb{R}^{d+1}$ of $\mathrm{epi}(g)$.  It is the least
upper semi-continuous, concave function satisfying $\mathrm{cl}(g)
\geq g$.  The function $g$ is \emph{closed} if $\mathrm{cl}(g) = g$.
An arbitrary function $h$ on $\mathbb{R}^d$ is \emph{continuous
relative} to a subset $S$ of $\mathbb{R}^d$ if its restriction to $S$
is a continuous function.  A non-zero vector $z \in \mathbb{R}^d$ is a
\emph{direction of increase} of $h$ on $\mathbb{R}^d$ if $t \mapsto
h(x+tz)$ is non-decreasing for every $x \in \mathbb{R}^d$.

The convex hull of finitely many points is called a \emph{polytope}.
The convex hull of $d+1$ affinely independent points is called a
$d$\emph{-dimensional simplex} (pl. \emph{simplices}).  If $C$ is a
convex set in $\mathbb{R}^d$, then a \emph{supporting half-space} to
$C$ is a closed half-space which contains $C$ and has a point of $C$
in its boundary.  A \emph{supporting hyperplane} $H$ to $C$ is a
hyperplane which is the boundary of a supporting half-space to $C$.
Thus $H = \{x \in \mathbb{R}^d: \langle x,b \rangle = \beta\}$, for
some $b \in \mathbb{R}^d$ and $\beta \in \mathbb{R}$ such that
$\langle x,b \rangle \leq \beta$ for all $x \in C$ with equality for
at least one $x \in C$.

If $V$ is a finite set of points in $\mathbb{R}^d$ such that $P =
\mathrm{conv}(V)$ is a $d$-dimensional polytope in $\mathbb{R}^d$,
then a \emph{face} of $P$ is a set of the form $P \cap H$, where $H$
is a supporting hyperplane to $P$.  The \emph{vertex set} of $P$,
denoted $\mathrm{vert}(P)$, is the set of $0$-dimensional faces
(\emph{vertices}) of $P$.  A \emph{subdivision} of $P$ is a finite set
of $d$-dimensional polytopes $\{S_1,\ldots,S_t\}$ such that $P$ is the
union of $S_1,\ldots,S_t$ and the intersection of any two distinct
polytopes in the subdivision is a face of both of them.  If $S =
\{S_1,\ldots,S_t\}$ and $\tilde{S} =
\{\tilde{S}_1,\ldots,\tilde{S}_{t'}\}$ are two subdivisions of $P$,
then $\tilde{S}$ is a \emph{refinement} of $S$ if each $S_l$ is
contained in some $\tilde{S}_{l'}$.  The \emph{trivial subdivision} of
$P$ is $\{P\}$.  A \emph{triangulation} of $P$ is a subdivision of $P$
in which each polytope is a simplex.

If $P$ is a $d$-dimensional polytope in $\mathbb{R}^d$, $F$ is a
$(d-1)$-dimensional face of $P$ and $v \in \mathbb{R}^d$, then there
is a unique supporting hyperplane $H$ to $P$ containing $F$.  The
polytope $P$ is contained in exactly one of the closed half-spaces
determined by $H$, and if $v$ is in the opposite open half-space, then
$F$ is \emph{visible} from $v$.  If $V$ is a finite set in
$\mathbb{R}^d$ such that $P = \mathrm{conv}(V)$, if $v \in V$ and $S =
\{S_1,\ldots,S_t\}$ is a subdivision of $P$, then the result of
\emph{pushing} $v$ is the subdivision $\tilde{S}$ of $P$ obtained by
modifying each $S_l \in S$ as follows:
\begin{enumerate}
\item If $v \notin S_l$, then $S_l \in \tilde{S}$
\item If $v \in S_l$ and $\mathrm{conv}(\mathrm{vert}(S_l) \setminus
\{v\})$ is $(d-1)$-dimensional, then $S_l \in \tilde{S}$
\item If $v \in S_l$ and $S_l' = \mathrm{conv}(\mathrm{vert}(S_l)
\setminus \{v\})$ is $d$-dimensional, then $S_l' \in \tilde{S}$.
Also, if $F$ is any $(d-1)$-dimensional face of $S_l'$ that is visible
from $v$, then $\mathrm{conv}(F \cup \{v\}) \in \tilde{S}$.
\end{enumerate}

If $\sigma$ is a convex function on $\mathbb{R}^n$, then $y' \in
\mathbb{R}^n$ is a \emph{subgradient} of $\sigma$ at $y$ if
\[ \sigma(z) \geq \sigma(y) + \langle y', z-y \rangle
\] for all $z \in \mathbb{R}^n$.  If $\sigma$ is differentiable at
$y$, then $\nabla \sigma(y)$ is the unique subgradient to $\sigma$ at
$y$; otherwise the set of subgradients at $y$ has more than one
element.  The \emph{one-sided directional derivative} of $\sigma$ at
$y$ with respect to $z \in \mathbb{R}^n$ is
\[ \sigma'(y;z) = \lim_{t \searrow 0} \frac{\sigma(y + tz) -
\sigma(y)}{t},
\] which always exists (allowing $-\infty$ and $\infty$ as limits)
provided $\sigma(y)$ is finite.

\section{Proofs}
\label{Sec:Proofs}
\begin{prooftitle}{\textsc{of Proposition~\ref{Prop:LCProperties}}}
(a) If $f$ is log-concave, then for $x \in \mathbb{R}^d$, we can write
\[ f_{X|P_V(X)}(x|t) \propto f(x)\mathbbm{1}_{\{P_V(x) = t\}},
\] a product of log-concave functions.  Thus $f_{X|P_V(X)}(\cdot|t)$
is log-concave for each $t$.

(b) Let $x_1,x_2 \in \mathbb{R}^d$ be distinct and let $\lambda \in
(0,1)$.  Let $V$ be the $(d-1)$-dimensional subspace of $\mathbb{R}^d$
whose orthogonal complement is \emph{parallel} to the \emph{affine
hull} of $\{x_1,x_2\}$ (i.e. the line through $x_1$ and $x_2$).
Writing $f_{P_V(X)}$ for the marginal density of $P_V(X)$ and $t$ for
the common value of $P_V(x_1)$ and $P_V(x_2)$, the density of $X$ at
$x \in \mathbb{R}^d$ is
\[ f(x) = f_{X|P_V(X)}(x|t)f_{P_V(X)}(t).
\] Thus $f$ is log-concave, as required.  \hfill $\Box$
\end{prooftitle}

\begin{completionprooftitle}{\textsc{of
Theorem~\ref{Thm:ExistUnique}}} We prove each of the steps (i)--(v)
outlined in Section~\ref{Sec:ExistUnique} in turn.  First note that if
$x_0 \in C_n$, then by Carath\'eodory's theorem (Theorem~17.1 of
\citet{Rockafellar1997}), there exist distinct indices
$i_1,\ldots,i_r$ with $r \leq d+1$, such that $x_0 = \sum_{l=1}^r
\lambda_l X_{i_l}$ with each $\lambda_l > 0$ and $\sum_{l=1}^r
\lambda_l = 1$.  Thus, if $f(x_0) = 0$, then by Jensen's inequality,
\[ -\infty = \log f(x_0) \geq \sum_{l=1}^r \lambda_l \log f(X_{i_l}),
\] so $f(X_i) = 0$ for some $i$.  But then $\psi_n(f) = -\infty$.
This proves~(i).

Now suppose $f(x_0) > 0$ for some $x_0 \notin C_n$.  Then $\{x:f(x) >
0\}$ is a convex set containing $C_n \cup \{x_0\}$, a set which has
strictly larger $d$-dimensional Lebesgue measure than that of $C_n$.
We therefore have $\psi_n(f) < \psi_n(f \mathbbm{1}_{C_n})$, which
proves~(ii).

To prove~(iii), we first show that $\log f$ is \emph{closed}.  Suppose
that $\log f(X_i) = y_i$ for $i=1,\ldots,n$ but that $\log f \neq
\bar{h}_y$.  Then since $\log f(x) \geq \bar{h}_y(x)$ for all $x \in
\mathbb{R}^d$, we may assume that there exists $x_0 \in C_n$ such that
$\log f(x_0) > \bar{h}_y(x_0)$.  If $x_0$ is in the \emph{relative
interior} of $C_n$, then since $\log f$ and $\bar{h}_y$ are continuous
at $x_0$ (by Theorem~10.1 of \citet{Rockafellar1997}), we must have
\[ \psi_n(f) < \psi_n\bigl(\exp(\bar{h}_y)\bigr).
\] The only remaining possibility is that $x_0$ is on the
\emph{relative boundary} of $C_n$.  But $\bar{h}_y$ is closed by
Corollary~17.2.1 of \citet{Rockafellar1997}, so writing
$\mathrm{cl}(g)$ for the \emph{closure} of a concave function $g$, we
have $\bar{h}_y = \mathrm{cl}(\bar{h}_y) = \mathrm{cl}(\log f) \geq
\log f$, where we have used Corollary~7.3.4 of \citet{Rockafellar1997}
to obtain the middle equality.  It follows that $\log f$ is closed and
$\log f = \bar{h}_y$, which proves~(iii).

Note that $\log f$ has no \emph{direction of increase}, because if $x
\in C_n$, $z$ is a non-zero vector and $t>0$ is large enough that
$x+tz \notin C_n$, then $-\infty = \log f(x+tz) < \log f(x)$.  It
follows by Theorem~27.2 of \citet{Rockafellar1997} that the supremum
of $f$ is finite (and is attained).  Using properties~(i) and~(ii) as
well, we may write $\int f(x) \, dx = c$, say, where $c \in
(0,\infty)$.  Thus $f(x) = c\bar{f}(x)$, for some $\bar{f} \in
\mathcal{F}_0$.  But then
\[ \psi_n(\bar{f}) - \psi_n(f) = -1 - \log c + c \geq 0,
\] with equality only if $c=1$.  This proves~(iv).

To prove~(v), we may assume by~(iv) that $\exp (\bar{h}_y)$ is a
density.  Let $\max_i \bar{h}_y(X_i) = M$ and let $\min_i
\bar{h}_y(X_i) = m$.  We show that when $M$ is large, in order for
$\exp (\bar{h}_y)$ to be a density, $m$ must be negative with $|m|$ so
large that $\psi_n\bigl(\exp(\bar{h}_y)\bigr) \leq \psi_n(f)$.  First
observe that if $x \in C_n$ and $\bar{h}_y(X_i) = M$, then for $M$
sufficiently large we must have $M-m > 1$, and then
\begin{align*} \bar{h}_y\Bigl(X_i + \frac{1}{M-m}(x-X_i)\Bigr) &\geq
\frac{1}{M-m} \bar{h}_y(x) + \frac{M-m-1}{M-m}\bar{h}_y(X_i) \\ &\geq
\frac{m}{M-m} + \frac{(M-m-1)M}{M-m} = M-1.
\end{align*} (The fact that $\bar{h}_y(x) \geq m$ follows by Jensen's
inequality.)  Hence, denoting Lebesgue measure on $\mathbb{R}^d$ by
$\mu$, we have
\[ \mu(\{x:\bar{h}_y(x) \geq M-1\}) \geq \mu\Bigl(\Bigl\{X_i +
\frac{1}{M-m}(C_n-X_i)\Bigr\}\Bigr) = \frac{\mu(C_n)}{(M-m)^d}.
\] Thus
\[ \int_{\mathbb{R}^d} \exp\{\bar{h}_y(x)\} \, dx \geq
e^{M-1}\frac{\mu(C_n)}{(M-m)^d}.
\] For $\exp (\bar{h}_y)$ to be a density, then, we require $m \leq -
\frac{1}{2}e^{(M-1)/d}\mu(C_n)^{1/d}$ when $M$ is large.  But then
\[ \psi_n\bigl(\exp (\bar{h}_y)\bigr) \leq \frac{(n-1)M}{n} -
\frac{1}{2n}e^{(M-1)/d}\mu(C_n)^{1/d} \leq \psi_n(f)
\] when $M$ is sufficiently large.  This proves~(v).

It is not hard to see that for any $M > 0$, the function $y \mapsto
\psi_n(\exp(\bar{h}_y)\bigr)$ is continuous on the compact set
$[-M,M]^n$, and thus the proof of the existence of a maximum
likelihood estimator is complete.  To prove uniqueness, suppose that
$f_1, f_2 \in \mathcal{F}$ and both $f_1$ and $f_2$ maximise
$\psi_n(f)$.  We may assume $f_1, f_2 \in \mathcal{F}_0$, $\log f_1,
\log f_2 \in \mathcal{H}$ and $f_1$ and $f_2$ are supported on $C_n$.
Then the normalised geometric mean
\[ g(x) = \frac{\{f_1(x)f_2(x)\}^{1/2}}{\int_{C_n}
\{f_1(y)f_2(y)\}^{1/2} \, dy},
\] is a log-concave density, with
\begin{align*} \psi_n(g) &= \frac{1}{2n}\sum_{i=1}^n \log f_1(X_i) +
\frac{1}{2n}\sum_{i=1}^n \log f_2(X_i) - \log \int_{C_n}
\{f_1(y)f_2(y)\}^{1/2} \, dy - 1 \\ &= \psi_n(f_1) - \log \int_{C_n}
\{f_1(y)f_2(y)\}^{1/2} \, dy.
\end{align*} However, by Cauchy--Schwarz, $\int_{C_n}
\{f_1(y)f_2(y)\}^{1/2} \, dy \leq 1$, so $\psi_n(g) \geq \psi_n(f_1)$.
Equality is obtained if and only if $f_1 = f_2$ almost everywhere, but
since $f_1$ and $f_2$ are \emph{continuous relative} to $C_n$
(Theorem~10.2 of \citet{Rockafellar1997}), this implies that
$f_1=f_2$.  An alternative way of proving the uniqueness of the
maximum likelihood estimator may be based on the fact that
$\psi_n\bigl(tf_1 + (1-t)f_2\bigr) > t\psi_n(f_1) + (1-t)\psi_n(f_2)$
for all $t \in (0,1)$, provided $f_1$ and $f_2$ are distinct elements
of $\mathcal{F}$.  \hfill $\Box$
\end{completionprooftitle}

\begin{prooftitle}{\textsc{of Theorem~\ref{Thm:sigma}}} For $t \in
(0,1)$ and $y^{(1)},y^{(2)} \in \mathbb{R}^n$, the function
$\bar{h}_{ty^{(1)}+(1-t)y^{(2)}}$ is the least concave function
satisfying $\bar{h}_{ty^{(1)}+(1-t)y^{(2)}}(X_i) \geq ty_i^{(1)} +
(1-t)y_i^{(2)}$ for $i=1,\ldots,n$, so
$\bar{h}_{ty^{(1)}+(1-t)y^{(2)}} \leq t\bar{h}_{y^{(1)}} +
(1-t)\bar{h}_{y^{(2)}}$.  The convexity of $\sigma$ follows from this
and the convexity of the exponential function.  It is clear that
$\sigma \geq \tau$, since $\bar{h}_y(X_i) \geq y_i$ for
$i=1,\ldots,n$.

From Theorem~\ref{Thm:ExistUnique}, we can find $y^* \in \mathbb{R}^n$
such that $\log \hat{f}_n = \bar{h}_{y^*}$ with $\bar{h}_{y^*}(X_i) =
y_i^*$ for $i=1,\ldots,n$, and this $y^*$ minimises $\tau$.  For any
other $y \in \mathbb{R}^n$ which minimises $\tau$, by the uniqueness
part of Theorem~\ref{Thm:ExistUnique} we must have $\bar{h}_y =
\bar{h}_{y^*}$, so $\sigma(y) > \sigma(y^*) = \tau(y^*)$.  \hfill
$\Box$
\end{prooftitle}

\subsection{Non-differentiability of \texorpdfstring{$\sigma$}{sigma}
  and computation of subgradients}

In this section, we find explicitly the set of points at which the
function $\sigma$ defined in~(\ref{Eq:sigmadef}) is differentiable,
and compute a subgradient of $\sigma$ at each point.  For
$i=1,\ldots,n$, define
\[ J_i = \{j = (j_1,\ldots,j_{d+1}) \in J: i = j_l \ \text{for some
$l=1,\ldots,d+1$}\}.
\] The set $J_i$ is the index set of those simplices $C_{n,j}$ that
have $X_i$ as a vertex.  Let $\mathcal{Y}$ denote the set of vectors $y =
(y_1,\ldots,y_n) \in \mathbb{R}^n$ with the property that for each $j
= (j_1,\ldots,j_{d+1}) \in J$, if $i \neq j_l$ for any $l$ then
\[
\bigl\{(X_i,y_i),(X_{j_1},y_{j_1}),\ldots,(X_{j_{d+1}},y_{j_{d+1}})\bigr\}
\] is affinely independent in $\mathbb{R}^{d+1}$.  This is the set of
points for which no tent pole is touching but not critically
supporting the tent.  Notice that the complement of $\mathcal{Y}$ has
zero Lebesgue measure in $\mathbb{R}^n$.  For $y \in \mathbb{R}^n$ and
$i=1,\ldots,n$, and in the notation of Section \ref{Sec:Computation},
let
\begin{equation}
\label{Eq:Subgrad}
\partial_i(y) = - \frac{1}{n} + \sum_{j \in J_i} |\det A_j| 
  \int_{T_d} e^{\langle w,z_j \rangle + y_{j_1}} \biggl\{\Bigl(1 - \sum_{l=1}^d
    w_l\Bigr)\mathbbm{1}_{\{j_1 = i\}} + 
  \sum_{l=1}^d w_l \mathbbm{1}_{\{j_{l+1} = i\}}\biggr\} \, dw . \notag
\end{equation}
\begin{prop}
\label{Prop:Partial} Assume \textbf{(A1)}.  (a) For $y \in
\mathcal{Y}$, the function $\sigma$ is differentiable at $y$ and for
$i=1,\ldots,n$ satisfies
\[ \frac{\partial \sigma}{\partial y_i}(y) = \partial_i(y).
\] (b) For $y \in \mathcal{Y}^c$, the function $\sigma$ is not
differentiable at $y$, but the vector
$(\partial_1(y),\ldots,\partial_n(y))$ is a subgradient of $\sigma$ at
$y$.
\end{prop}
\begin{proof} By Theorem~25.2 of \citet{Rockafellar1997}, it suffices
to show that for $y \in \mathcal{Y}$, all of the partial derivatives
exist and are given by the expression in the statement of the
proposition.  For $i=1,\ldots,n$ and $t \in \mathbb{R}$, let $y^{(t)}
= y + te_i^n$, where $e_i^n$ denotes the $i$th unit coordinate vector in 
$\mathbb{R}^n$.  For sufficiently small values of $|t|$, we may write
\[ \bar{h}_{y^{(t)}}(x) = \left\{ \begin{array}{ll} \langle
x,b_j^{(t)} \rangle - \beta_j^{(t)} & \mbox{\text{if $x \in C_{n,j}$
for some $j \in J$}} \\ -\infty & \mbox{if $x \notin
C_n$,} \end{array} \right.
\] for certain values of $b_1^{(t)},\ldots,b_m^{(t)} \in \mathbb{R}^d$
and $\beta_1^{(t)},\ldots,\beta_m^{(t)} \in \mathbb{R}$.  If $j \notin
J_i$, then $b_j^{(t)} = b_j$ and $\beta_j^{(t)} = \beta_j$ for
sufficiently small $|t|$.  On the other hand, if $j \in J_i$, then there
are two cases to consider:
\begin{enumerate}
\item If $j_1  = i$, then for sufficiently small $t$, we have
  $z_j^{(t)} = z_j - t1_d$, where $1_d$ denotes a $d$-vector of ones,
  so that $b^{(t)}_j = b_j - t(A_j^T)^{-1}1_d$ and $\beta^{(t)}_j =
  \beta_j - t (1 + \langle A_j^{-1} \alpha_j, 1_d \rangle)$
\item If $j_{l+1} = i$ for some $l \in \{1, \ldots , d\}$, then for
  sufficiently small $t$, we have $z_j^{(t)} = z_j + te_l^d$, so that
  $b^{(t)}_j = b_j + t(A_j^T)^{-1}e_l^d$ and $\beta_j^{(t)} = \beta_j
  + t \langle A_j^{-1} \alpha_j, e_l^d\rangle$.
\end{enumerate} 
It follows that
\begin{equation}
\label{eq:partial}\begin{split}
  \frac{\partial \sigma}{\partial y_i}(y) &= -\frac{1}{n} + \lim_{t
    \rightarrow 0} \frac{1}{t} \sum_{j \in J_i} \int_{C_{n,j}}
  \exp\bigl\{\langle x , b_j^{(t)}\rangle - \beta_j^{(t)}
  \bigr\} - \exp\left\{ \langle x , b_j \rangle - \beta_j\right\}  \, dx \notag \\
  &= -\frac{1}{n} + \lim_{t \rightarrow 0} \frac{1}{t} \sum_{j \in
    J_i} \left[ \int_{C_{n,j}} e^{ \langle x , b_j\rangle - \beta_j}
    \bigl\{e^{t  (1 - \langle A_j^{-1}(x-\alpha_j), 1_d\rangle)} - 1\bigr\} \, dx\mathbbm{1}_{\{j_1 = i\}} \right. \notag \\
  &\hspace{5cm}+ \left. \sum_{l=1}^d \int_{C_{n,j}} e^{ \langle x, b_j \rangle -
      \beta_j}\bigl\{e^{t \langle A_j^{-1}( x-\alpha_j), e_l^d\rangle } - 1 \bigr\}\, dx\mathbbm{1}_{\{j_{l+1} = i\}} \right] \notag\\
  &= \partial_i(y),
\end{split}
\end{equation}
where to obtain the final line we have made the substitution $x = A_j
w + \alpha_j$, after taking the limit as $t \rightarrow 0$.

(b) If $y \in \mathcal{Y}^c$, then it can be shown that there exists a
unit coordinate vector $e_i^n$ in $\mathbb{R}^n$ such that the
\emph{one-sided directional derivative} at $y$ with respect to $e_i^n$,
denoted $\sigma'(y;e_i^n)$, satisfies $\sigma'(y;e_i^n) >
-\sigma'(y;-e_i^n)$.  Thus $\sigma$ is not differentiable at $y$.  To
show that $\partial(y) = (\partial_1(y),\ldots,\partial_n(y))$ is a
subgradient of $\sigma$ at $y$, it is enough by Theorem~25.6
of~\citet{Rockafellar1997} to find, for each $\epsilon > 0$, a point
$\tilde{y} \in \mathbb{R}^n$ such that $\|\tilde{y} - y\| < \epsilon$
and such that $\sigma$ is differentiable at $\tilde{y}$ with $\|\nabla
\sigma(\tilde{y}) - \partial(y)\| < \epsilon$.  This can be done by
sequentially making small adjustments to the components of $y$ in the
same order as that in which the vertices were \emph{pushed} in
constructing the triangulation.  \hfill $\Box$
\end{proof}

A subgradient of $\sigma$ at any $y \in \mathbb{R}^n$ may be computed
using Proposition~\ref{Prop:Partial}, (\ref{Eq:Subgrad}) and
(\ref{Eq:Int}) once we have a formula for
\[ \tilde{I}_{d,u}(z) = \int_{T_d} w_u \exp\biggl(\sum_{r=1}^d
z_rw_r\biggr) \, dw,
\] when $z_1,\ldots,z_d$ are non-zero and distinct.  In
\citet{CSS2008}, it is shown that the required formula is
\begin{align}
\label{Eq:Idtilde} \tilde{I}_{d,u}(z) &=
\sum_{\stackrel{\scriptstyle{1 \leq r \leq d}}{r \neq u}}
\frac{e^{z_r}}{z_r(z_r-z_{u})}\prod_{\stackrel{\scriptstyle{1 \leq s
\leq d}}{s \neq r}} \frac{1}{(z_r - z_s)} -
\sum_{\stackrel{\scriptstyle{1 \leq r \leq d}}{r \neq u}}
\frac{e^{z_{u}}}{z_r(z_r-z_{u})} \prod_{\stackrel{\scriptstyle{1 \leq
s \leq d}}{s \neq r}} \frac{1}{(z_r - z_s)} \nonumber \\
&\hspace{6.5cm} + \frac{(-1)^d(e^{z_u}-1)}{z_u \prod_{s=1}^d z_s} +
\frac{e^{z_u}}{z_u}\prod_{\stackrel{\scriptstyle{1 \leq s \leq d}}{s
\neq u}} \frac{1}{(z_u - z_s)}.
\end{align}

\bibliography{css}

\end{document}